\documentclass[journal]{IEEEtran}
\usepackage{algorithm}
\usepackage{algorithmic}
\usepackage{array}

\newtheorem{myLemma}{\textit{Lemma}}
\newtheorem{myRemark}{\textit{Remark}}
\usepackage{stfloats}
\usepackage{cite}
\usepackage{amsmath,amssymb,amsfonts,bm}
\allowdisplaybreaks
\usepackage{graphicx}
\usepackage{epstopdf}
\usepackage{subfigure} 
\usepackage[caption=false, font=normalsize, labelfont=sf, textfont=sf]{subfig}

\usepackage{textcomp}
\usepackage{stfloats}
\usepackage{url}
\usepackage{verbatim}
\usepackage{enumerate}
\usepackage{amssymb}
\usepackage{xcolor}
\def\BibTeX{{\rm B\kern-.05em{\sc i\kern-.025em b}\kern-.08em
    T\kern-.1667em\lower.7ex\hbox{E}\kern-.125emX}}
\usepackage{xpatch}
    \makeatletter
\def\changeBibColor#1{%
\in@{#1}{ }
\ifin@\color{blue}\else\normalcolor\fi
}
\xpatchcmd
\@bibitem
{\item}
{\changeBibColor{#1}\item}
{}{\fail}
\xpatchcmd\@lbibitem
{\item}
{\changeBibColor{#2}\item}
{}{\fail}
\makeatother

\begin{document}

\title{Beam Training for Pinching-Antenna Systems (PASS)}
\author{Suyu Lv,~\IEEEmembership{Member,~IEEE},	
 Yuanwei Liu,~\IEEEmembership{Fellow,~IEEE}
 and Zhiguo Ding,~\IEEEmembership{Fellow,~IEEE}
	\thanks{Suyu Lv is with School of Information Science and Technology, Beijing University of Technology, Beijing, 100124, China (e-mail: lvsuyu@bjut.edu.cn).}
	\thanks{Yuanwei Liu is with the Department of Electrical and Electronic Engineering, The University of Hong Kong, Pokfulam, 999077, Hong Kong (e-mail: yuanwei@hku.hk).}
	\thanks{Zhiguo Ding is with Khalifa University, Abu Dhabi, UAE, and the University of Manchester, Manchester, M1 9BB, UK (email: z.ding@lancaster.ac.uk).}
}

\maketitle

\begin{abstract}
	This article investigates the beam training design problems for pinching-antenna systems (PASS), where single-waveguide-single-user (SWSU), single-waveguide-multi-user (SWMU) and multi-waveguide-multi-user (MWMU) scenarios are considered. 
	For SWSU-PASS, we design a scalable codebook, based on which we propose a three-stage beam training (3SBT) scheme.
	Specifically, 1) firstly, the 3SBT scheme utilizes one activated pinching antenna to obtain a coarse one-dimensional location at the first stage; 2) secondly, it achieves further phase matching with an increased number of activated antennas at the second stage; 3) finally, it realizes precise beam alignment through an exhaustive search at the third stage. 
	For SWMU-PASS, based on the scalable codebook design, we propose an improved 3SBT scheme to support non-orthogonal multiple access (NOMA) transmission. 
	For MWMU-PASS, we first present a generalized  expression of the received signal based on the partially-connected hybrid beamforming structure.
	Furthermore, we introduce an increased-dimensional scalable codebook design, based on which an increased-dimensional 3SBT scheme is proposed. 
	Numerical results reveal that: i) the proposed beam training scheme can significantly reduce the training overhead compared to the two-dimensional exhaustive search, while maintaining reasonable rate performance; ii) compared to fixed-location pinching antennas and  conventional array antennas, the proposed dynamic pinching antennas yield better flexibility and improved performance.
\end{abstract}


\section{Introduction}

To support the higher frequencies and larger bandwidths required by the sixth generation mobile communications (6G), one of the primary focuses in 6G research is the development of novel antenna technologies, featured by their smaller volumes, larger scales, and increased flexibility and reconfigurability \cite{MVT.2022.3233157}. 
These advancements are crucial for achieving ultra-high transmission data rates and meeting stringent demands of future communication systems. 
Currently, the research on 6G novel antenna technologies encompasses various areas, including but not limited to millimeter-wave (mmWave) and terahertz (THz) antennas, reconfigurable intelligent surfaces (RIS) \cite{JSAC.2020.3007211, ACCESS.2019.2935192}, fluid antennas \cite{TWC.2023.3276245}, and movable antennas \cite{MCOM.001.2300212}.
Each of these technologies presents unique opportunities and challenges for the evolution of 6G.
For instance, while RIS-aided systems can enable active channel reconstruction and adjustment, they suffer from a double fading effect, and their passive nature makes it difficult to acquire channel state information (CSI).
Fluid antennas exhibit high flexibility, enabling rapid response to environmental changes, yet they face issues of insufficient robustness, and incapable to combat large-scale pathloss \cite{OJAP.2021.3069325}.

\begin{figure}[t]
	\centering
	\includegraphics[width=7.0cm]{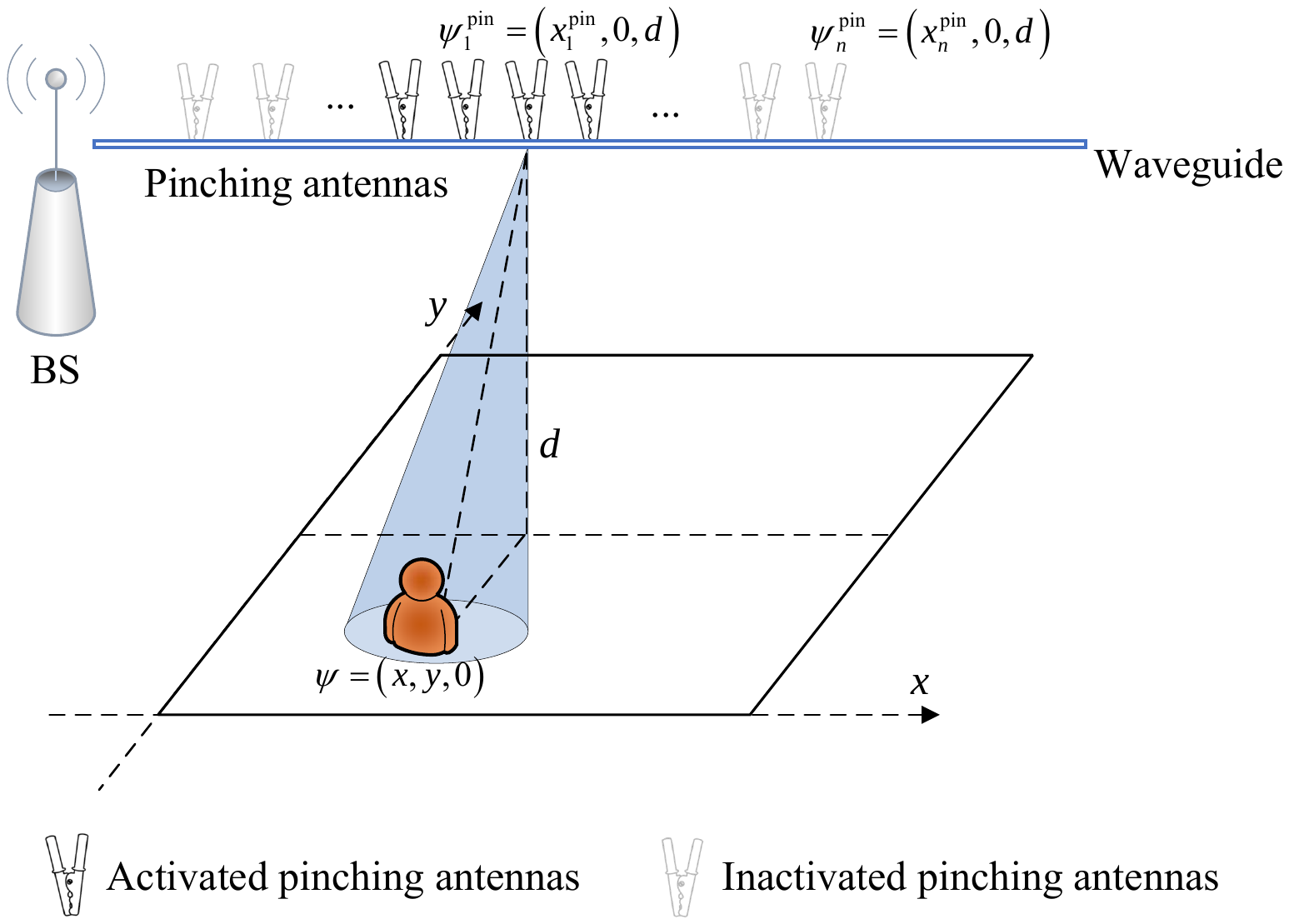}
	\caption{Illustration of PASS.}
	\label{SystemModel_SWSU}
\end{figure}

To tackle these issues, a novel flexible-antenna technology, termed the pinching antenna, has been proposed. 
This technology was first developed by DOCOMO in 2022 \cite{DOCOMO}, and has recently begun to gain attention from the academics \cite{arXiv:2501.18409, arXiv.2412.02376}. 
The concept of pinching antennas involves the activation of small dielectric particles, such as plastic pinches, on a dielectric waveguide, as illustrated in Fig. {\ref{SystemModel_SWSU}}. 
The pinching antenna offers several compelling advantages that make it an attractive solution for future wireless communication systems. 
Firstly, it can be deployed along an arbitrarily long waveguide, allowing for large-scale adjustment of the locations of activated antennas \cite{DOCOMO_PDF}. This capability enables the antennas to be situated very close to their target receivers, thereby establishing strong line-of-sight (LoS) links in the near-field communication (NFC) region. 
Secondly, it yields remarkable antenna reconfiguration. The scale of the pinching-antenna systems (PASS) can be easily adjusted by either activating more pinching antennas or deactivating existing ones on the waveguide, which realizes a dynamic and reconfigurable antenna system. Furthermore, multiple antennas can also be flexibly deployed on one or multiple waveguides, which paves a new way for the implementation of multiple-input multiple-output (MIMO) technology.
Lastly, pinching antennas are characterized by their extremely low costs, making PASS a highly cost-effective solution.

Despite these advantages of PASS, the acquisition of CSI for PASS faces unique challenges compared to conventional fixed-location antennas. 
On the one hand, the location of a pinching antenna needs to be dynamically changed in order to be tailored to its serving user's location, which leads to dynamic variations in the channel environment. On the other hand, due to the unpredictability of the locations of the users and their pinching antennas, it is difficult to leverage historical data to establish an accurate channel prediction model \cite{TWC.2022.3219052}.
We note that the locations of the activated antennas are closely related to the beam direction, which in turn affects the overall system performance. A crucial issue to be addressed is how to obtain users' CSI and dynamically adjust the antenna location, which is key to optimize the direction of the transmitted beam accordingly. 
This motivates us to explore near-field beam training solutions. 
Instead of the simplified plenary wave model, the more accurate spherical wave model is more appropriate to be used in PASS. Recall that with the spherical wave model, the CSI contains both angular and distance domains \cite{OJCOMS.2023.3305583}, resulting in a significant overhead for near-field beam training, especially in large-scale MIMO systems.
To achieve a balance between training performance and training overhead, \cite{GLOBECOM54140.2023.10437041} and \cite{TVT.2023.3311868} proposed hierarchical training schemes. Specifically, the scheme in  \cite{GLOBECOM54140.2023.10437041} gradually increases training resolution on a two-dimensional (2D) plane, while the scheme in \cite{TVT.2023.3311868} performs angular and distance estimation in two separated stages.
The authors in \cite{TWC.2022.3222198} proposed a fast beam training scheme for wideband near-field applications, achieving beam focusing at multiple locations at multiple frequencies.
The near-field beam training problems in RIS-assisted MIMO systems were investigated in \cite{TCOMM.2023.3278728, TWC.2024.3393412}. 
However, the existing near-field beam training schemes, whether implemented at base station (BS) or RIS, assume that the antenna array is in a fixed position, which are not applicable to PASS.

The research on pinching antennas is in its infancy stages. 
Despite the enhanced flexibility offered by PASS, the CSI is inherently coupled with the locations of the activated antennas, which in turn affects the user's received signal strength and phase. 
In this paper, we aim to acquire CSI while concurrently determining the optimal locations for the activated pinching antennas, thereby facilitating more efficient and reliable communication in PASS.
To achieve this objective, we propose beam training schemes for scenarios involving single-waveguide-single-user (SWSU), single-waveguide-multi-user (SWMU) and multi-waveguide-multi-user (MWMU) configurations, which avoid direct channel estimation and hence significantly reduce system overhead. 
The main contributions of our works can be summarized as
\begin{itemize}
	\item For SWSU-PASS, a scalable codebook generation design is introduced, based on which a three-stage beam training (3SBT) scheme is proposed. Specifically, at the first stage, one activated pinching antenna is utilized to obtain a coarse one-dimensional location; at the second stage, further phase matching is achieved in another dimension with an  increased number of activated antennas; at the third stage, precise beam alignment is realize through an exhaustive search. 
	\item For SWMU-PASS, an improved 3SBT scheme is proposed to support non-orthogonal multiple access (NOMA) transmission.
	Specifically, at the first stage, activated pinching antennas are grouped into several clusters for individual user training; at the second stage, the antennas are reclustered to avoid overlapping; at the third stage, a joint exhaustive search is conducted for multi-user beam training.
	\item For MWMU-PASS, a generalized  expression of the received signal is presented based on the partially-connected hybrid beamforming structure. An increased-dimensional scalable codebook design is introduce, based on which an increased-dimensional 3SBT scheme is proposed. Specifically, the waveguide for each user is determined at the first stage, where the separated user training and joint multi-user training are carried out in the second and third stages, respectively. 
	\item Numerical results demonstrate the effectiveness of our proposed beam training schemes for PASS. Compared to the 2D exhaustive search, the proposed schemes significantly reduced the training overhead while maintaining comparable rate performance. Additionally, the proposed pinching antenna scheme yield better flexibility and improved system performance compared to fixed-location pinching antennas and  conventional array antennas. 
\end{itemize}

The organization of this article is as follows. Sections II, III and IV focus on SWSU-PASS, SWSU-PASS and SWSU-PASS respectively, each encompassing the system model, the codebook generation design, and the beam training scheme. Simulation results and analysis, as well as conclusions, are presented in Sections V and VI, respectively.

\section{Single-Waveguide PASS Serving A Single User} \label{1WG_1U}

In this section, we focus on the scenario of SWSU-PASS. 
Firstly, we introduce the system model of SWSU-PASS. 
Subsequently, we present a scalable codebook generation design.
Finally, based on the predefined codebook, we propose a 3SBT scheme. 
The details are presented in the following subsections, respectively.

\subsection{System Model of SWSU-PASS}

The waveguide is assumed to be
placed parallel to the $x$-axis, and the height of the antenna is denoted by $d$. 
We consider an ideal scenario where the pinching antennas can be activated perfectly at any desirable locations along the waveguide. In other words, pinching antennas are continuously deployed along the waveguide. 
Without loss of generality, assume that $N$ pinching antennas are activated on a single waveguide{\footnote{Although the number of pinching antennas in the considered ideal scenario is potentially infinite, we assume that the number of antennas that can be activated at the same time is limited. In practice, the number of antennas can be optimized based on the system's available resources, hardware constraints, and the users' quality of service (QoS) requirements.}}, where the $n$-th pinching antenna is denoted by ${\rm{Pin}}_n$. The {\bf{general notation}} for the location of ${\rm{Pin}}_n$ is $\tilde \psi _n^{\rm{pin}} = \left(x_n^{\rm{pin}}, 0, d\right)$, as shown in Fig. {\ref{SystemModel_SWSU}}.
The spherical wave channel model for NFC is used \cite{OJCOMS.2023.3305583}. 
Recall that the pinching antennas can be positioned close to the user, allowing for the establishment of LoS links.
Therefore, for a user located at ${\psi} = \left(x, y, 0\right)$, the channel between the pinching antennas and the user is  \cite{arXiv.2412.02376}
\begin{equation}
	\begin{array}{l}
	{{\bf{h}}} = {\left[ {\frac{{\sqrt \eta  {e^{ - j\frac{{2\pi }}{\lambda }\left(\left| {{\psi} - \tilde \psi _1^{{\rm{pin}}}} \right| + \theta_1 \right)}}}}{{\left| {{\psi} - \tilde \psi _1^{{\rm{pin}}}} \right|}} \cdots \frac{{\sqrt \eta  {e^{ - j\frac{{2\pi }}{\lambda }\left(\left| {{\psi} - \tilde \psi _N^{{\rm{pin}}}} \right|+\theta_N \right)}}}}{{\left| {{\psi} - \tilde \psi _N^{{\rm{pin}}}} \right|}}} \right]^T},
	\end{array}
\end{equation}
where $\lambda$ is the wavelength, and $\eta  = \frac{{{c^2}}}{{16{\pi ^2}f_c^2}}$ with $c$ denoting the speed of light and $f_c$ denoting the carrier frequency. 
Note that the phase shifts $e^{-j \theta_n}$ are due to the signals propagating through the waveguide, and the phase shifts $e^{ - j\left( \frac{{2\pi }}{\lambda } \left| {\psi  - \tilde \psi _n^{{\rm{pin}}}} \right| \right) }$ are due to signals' free-space propagation. 
${\theta _n}$ denotes the phase shift experienced at ${\rm{Pin}}_n$, which is a function of the location of this antenna, e.g., $\theta_n = \frac{2\pi}{\lambda_g }\left| \psi_0^{\rm{pin}} -\tilde\psi_n^{\rm{pin}}\right| $, with $\psi_0^{\rm{pin}}$ denoting the location of the feed point of the waveguide, and $\lambda_g$ denoting the wavelength in the waveguide, $\lambda_g = \frac{\lambda}{n_{\rm{eff}}}$.

Since the pinching antennas are  located on the same waveguide, the signal transmitted by one pinching antenna is a phase-shifted version of the signal transmitted by another, which means that only one data stream can be supported.
For the convenience of analysis, we consider the total transmit power $P$ to be equally shared among $N$ activated pinching antennas. 
Thus, the pinching beamforming vector is given by ${\bf{g}} = \left[\sqrt{\frac{P}{N}}, \cdots, \sqrt{\frac{P}{N}} \right] \in {\mathbb{C}}^{N \times 1}$.
By treating the $N$ activated antennas as elements of a conventional linear array, the signal received by the user can be expressed as follows \cite{TIT.2012.2191700}
\begin{equation}
	\begin{array}{l}
	v = {\bf{h}}^H{\bf{g}}s + w 
	=  \left( {\sum\limits_{n = 1}^N {\frac{{\sqrt \eta  {e^{ - j\left( \frac{{2\pi }}{\lambda } \left| {\psi  - \tilde \psi _n^{{\rm{pin}}}} \right| + {\theta _n} \right) }}}}{{\left| {\psi  - \tilde \psi _n^{{\rm{pin}}}} \right|}}}} \right)\sqrt {\frac{P}{N}} s + w,
	\end{array}
\end{equation}
where $s$ is the signal passed along the waveguide, and $w\sim {\cal{CN}} \left(0, \sigma^2 \right)$ denotes the additive white Gaussian noise (AWGN).

\subsection{Scalable Codebook Generation} \label{Codebook}
 
Since users are located on the ground, we only need to consider codebook design and beam training schemes in a 2D plane.
Denote the sampling range by ${\cal{D}}$.
The {\bf{specific location}} on the waveguide that is closest to the given point ${\psi_f} = \left(x_f, y_f, 0\right)$ is denoted by ${\psi_f^{\rm{pin}}} = \left(x_f, 0, d\right)$. 
When the user located at ${\psi_f}$ is served, it is ideal to place all pinching antennas as close to ${\psi_f^{\rm{pin}}}$ as possible.
Recall that moving an antenna a few wavelengths for satisfying $\frac{{2\pi }}{\lambda }\left| {{\psi_f} - \tilde \psi _n^{{\rm{pin}}}} \right| + {\theta _n} = 2k\pi $ has a limited impact on the distance $\left| {{\psi_f} - \tilde \psi _n^{{\rm{pin}}}} \right|$ as well as the free-space pathloss. 
Therefore, we expect that all $N$ pinching antennas are clustering around ${\psi_f^{\rm{pin}}}$.

Based on the above discussions, we propose a scalable codebook design{\footnote{The codebook is stored in the form of a matrix, where each column represents a codeword. Unlike traditional codewords which represent the weighting vectors for beamforming, the codewords mentioned in this paper refer to the locations of the activated pinching antennas.}}.
For a given set of sampling points ${\cal{F}}$ and the number of activated pinching antennas $ \tilde N$, where $\tilde N$ is a general notation referring to the number of activated pinching antennas, the corresponding codebook can be generated through the following steps:
\begin{enumerate}[S1.]
	\item \label{S1} With given sampling point located at ${\psi_f}= \left(x_f, y_f, 0\right)  \in {\cal{D}}$, the location of the first activated pinching antenna is obtained by focusing on the segment between ${\psi_f^{\rm{pin}}}$ and the end of the waveguide and using the first-found location satisfying $
	\bmod \left\{ {\frac{{2\pi }}{\lambda }\left| {{\psi_f} - \tilde \psi _1^{{\rm{pin}}}} \right| + {\theta _1}, 2\pi } \right\} = 0 $, where
	$ \bmod \left\{ a, b \right\} $ is the modulo operation of $a$ by $b$.
	\item \label{S2} The location of the second activated pinching antenna can be obtained by focusing on the segment between $ \tilde \psi _{1}^{{\rm{pin}}} - \tilde \Delta$ and the feed point of the waveguide and using the first-found location satisfying
	$\bmod \left\{ {\frac{{2\pi }}{\lambda }\left| {{\psi_f} - \tilde \psi _2^{{\rm{pin}}}} \right| + {\theta _2}, 2\pi } \right\} = 0$, where $\tilde \Delta$ is the guard distance to avoid antenna coupling.
	\item \label{S3} Successively, the location of the $n$-th ($n$ is an odd number) activated pinching antenna, denoted by $ \tilde \psi _{N_{\rm{odd}}}^{{\rm{pin}}}$, can be obtained by focusing on the segment between $ \tilde \psi _{n-2}^{{\rm{pin}}} + \tilde \Delta$ and the end of the waveguide and using the first-found location which satisfies
	$\bmod \left\{ {\frac{{2\pi }}{\lambda }\left| {{\psi_f} - \tilde \psi _n^{{\rm{pin}}}} \right| + {\theta _n},2\pi } \right\} = 0$.
	\item \label{S4} The location of the $n$-th ($n$ is an even number) activated pinching antenna, denoted by $ \tilde \psi _{N_{\rm{even}}}^{{\rm{pin}}}$, can be obtained by focusing on the segment between $ \tilde \psi _{n-2}^{{\rm{pin}}} - \tilde \Delta$ and the feed point of the waveguide and using the first-found location which satisfies
	$\bmod \left\{ {\frac{{2\pi }}{\lambda }\left| {{\psi_f} - \tilde \psi _n^{{\rm{pin}}}} \right| + {\theta _n},2\pi } \right\} = 0$.
	\item \label{S5} Repeat the above process until $n= \tilde N$. Then store the generated codeword ${\bf{g}}_f$ in the codebook ${\cal{G}}$. 
\end{enumerate}

Through the above steps S{\ref{S1}} to S{\ref{S5}}, an initial codebook can be generated for a given number of sampling points $F = \left|{\cal{F}}\right|$ and a number of activated pinching antennas $\tilde N$.
It is worth noting that, when the waveguide location is fixed, for a user located at the sampling point ${\psi_f}$, the corresponding optimal locations for the activated pinching antennas are stored in codeword ${\bf{g}}_f$. 
The proposed codebook design is scalable:
\begin{itemize}
	\item When new sampling points are added, the corresponding codewords for these new sampling points can be added to the existing codebook by following steps S{\ref{S1}} to S{\ref{S5}}. 
	\item When the number of activated pinching antennas increases, the optimal locations for the first $\tilde N$ antennas can still be obtained from the existing codebook. It only requires adding the locations of the activated antennas starting from the $\left({\tilde N}+1\right)$-th one to each codeword. 
	\item If the number of activated pinching antennas, $N'$, is less than the number of pinching antennas stored in the codebook, i.e., $N' < \tilde N$, only the first $N'$ entries of the codeword vector need to be taken, as the antennas listed earlier are closer to ${\psi_f^{\rm{pin}}}$.
\end{itemize}

The proposed scalable codebook generation scheme is summarized in {\bf{Algorithm \ref{alg:1}}}.
Based on this pre-designed codebook, in the following subsection, we will present a beam training scheme.

\begin{algorithm}[!t]
	\caption{Scalable codebook generation scheme}
	\label{alg:1}
	\begin{algorithmic}[1]
		\begin{small}
		\REQUIRE set of sampling points ${\cal{F}}$, set of pinching antennas ${\cal{N}}$;
		\ENSURE codebook ${\cal{G}}$;
		\STATE {\bf{Generation of the initial codebook:}}
		\FORALL{sampling points ${f} \in {\cal{F}}$}
		\STATE  S{\ref{S1}} -- S{\ref{S5}};
		\ENDFOR
		\STATE {\bf{Extension of the number of sampling points:}} 
		\FORALL{newly added sampling points ${f_{\rm{new}}} \in {\cal{F}}_{\rm{new}}$}
		\STATE  S{\ref{S1}} -- S{\ref{S5}};
		\STATE  ${\cal{F}} \gets {\cal{F}} \cup f_{\rm{new}}$;
		\STATE Update codebook ${\cal{G}}$;
		\ENDFOR		
		\STATE {\bf{Extension of the number of pinching antennas:}}
		\STATE Renumber the newly added antennas starting from $N+1$;
		\FORALL{newly added pinching antennas ${n_{\rm{new}}} \in {\cal{N}}_{\rm{new}}$}		
			\FORALL{sampling points ${f} \in {\cal{F}}$}
			\STATE  S{\ref{S3}} -- S{\ref{S5}};
			\ENDFOR
			\STATE  ${\cal{N}} \gets {\cal{N}} \cup {\rm{Pin}}_{n_{\rm{new}}}$;
			\STATE Update codebook ${\cal{G}}$;
		\ENDFOR
		\end{small}
	\end{algorithmic}
\end{algorithm}

\subsection{3SBT scheme} \label{Beamtraining_Single}

To find the optimal codeword for a single user, the most straightforward approach is to exhaustively search through all codewords in the codebook. However, the cost of an exhaustive search can be significant when the number of sampling points or activated pinching antennas is large, while the beam training performance may suffer when the number of sampling points or activated pinching antennas is small.
Therefore, we are dedicated to proposing an efficient beam training scheme that can reduce the training overhead and guarantee training accuracy simultaneously.

\begin{figure}[t!]
	\centering
	\includegraphics[width=7.0cm]{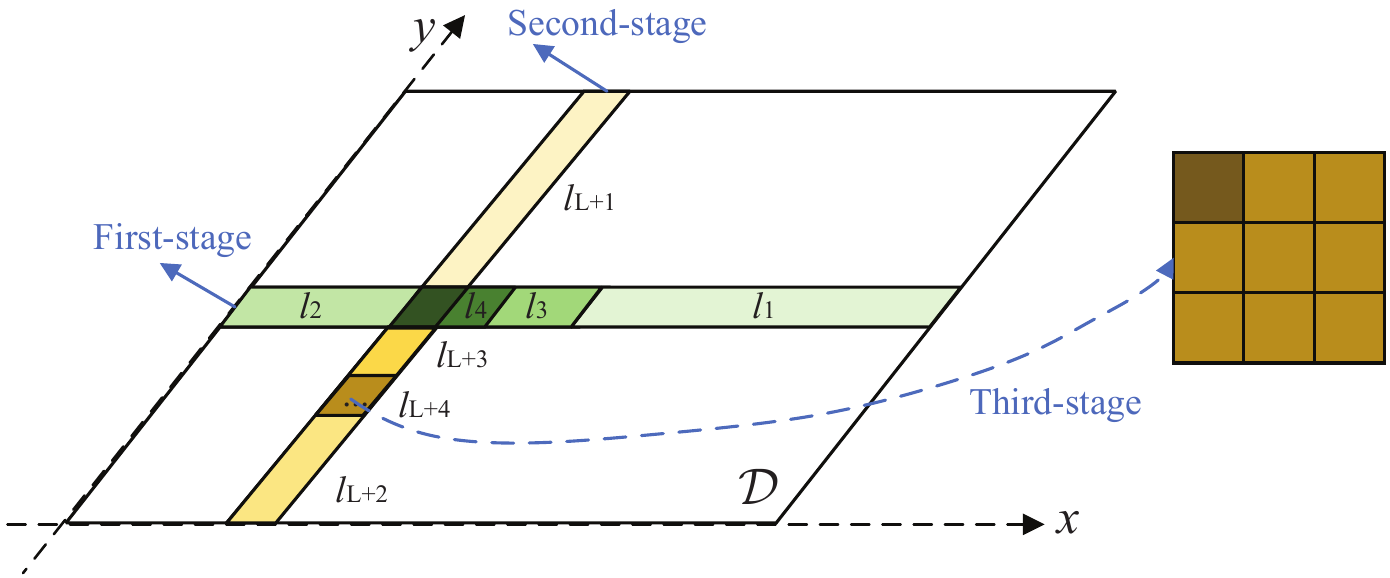}
	\caption{Illustration for the three-stage beam training scheme.}
	\label{BeamTraining_ThreeStage}
\end{figure}

Since the activated pinching antennas located on the waveguide are parallel to the $x$-axis, we can first perform beam training along the $x$-axis dimension. 
In this way, only one-dimensional sampling is required. Assume that there are $K$ sampling points along the one-dimensional axis. To achieve the same sampling accuracy, $K^2$ sampling points are required if 2D sampling is used, which significantly increases the training overhead. 
Based on this discussion, we propose a 3SBT scheme, as shown in Fig. {\ref{BeamTraining_ThreeStage}}.

\subsubsection{Stage 1 - Coarse-grained estimation} 
At the first stage, we aim to roughly locate the user's location along the $x$-axis dimension with a small amount of overhead. 
Since variations in the antenna's position along the $x$-axis dimension influence free-space pathloss, the received signal strength can be used to measure the distance between the user and the antenna. Therefore, there is no need for beam directionality, which means that a rough estimation of the user's location along the $x$-axis dimension can be obtained with only one antenna.

However, there exists an inevitable tradeoff between sampling accuracy and searching overhead. 
The performance will be better with a higher sampling accuracy, i.e., a smaller sampling space, but the searching overhead will be higher. 
Conversely, the performance will be degraded with a lower sampling accuracy, i.e., a larger sampling space, but the searching overhead will be smaller. 
Therefore, we design a hierarchical beam training scheme to gradually reduce the sampling space.
In each layer, the sampling range is divided into $K$ equal parts along the $x$-axis dimension.
Since we focus on the $x$-axis dimension at the first stage, we denote the sampling range by ${\cal{D}} = \left[x_{\rm{min}}, x_{\rm{max}}\right]$, and denote the $k$-th sub-range in $l$-th layer by ${\cal{D}}_{l,k} = \left[x_{\rm{min}}^{l,k}, x_{\rm{max}}^{l,k}\right]${\footnote{Although the sampling area is represented in one dimension here, it actually remains on a 2D plane, where the $y$-coordinate remains unchanged. This is merely to simplify the notation.}}.
The midpoint of ${\cal{D}}_{l,k}$ is denoted by $\psi_{l,k} = \left(x_{l,k}, 0, 0\right)$ with $x_{l,k} = \left(x_{\rm{min}}^{l,k}+x_{\rm{max}}^{l,k}\right)/2$.
The optimal sub-range ${\cal{D}}_{l,k}^{\rm{opt}}$ for each layer will serve as the sampling range for the next layer. 
That is, in the $\left(l+1\right)$-th layer, the sampling range ${\cal{D}}$ will be updated by ${\cal{D}}_{l,k}^{\rm{opt}}$.

The initial sampling range is ${\cal{D}} = \left[0,  D_x\right]$. 
In the $l$-th layer, ${\cal{D}}$ is divided into $K$ equal parts with length $\frac{D_x}{K^l}$. 
The $k$-th sub-range in $l$-th layer can be expressed as
\begin{equation}
	\begin{array}{l}
	{\cal{D}}_{l,k}  =\left[x_{\rm{min}}^{l,k},  x_{\rm{max}}^{l,k} \right] = \left[x_{\rm{min}}^{l-1,k} + \frac{\left(k-1\right)D_x}{K^l},  x_{\rm{min}}^{l-1,k} + \frac{kD_x}{K^l}\right], 
	\end{array}
\end{equation}
where the midpoint is
\begin{equation}
	\begin{array}{l}
	\psi_{l,k} = \left(x_{\rm{min}}^{l-1,k} + \frac{kD_x}{K^l} - \frac{D_x}{2K^l}, 0, 0\right),
	\end{array}
\end{equation}
which is the sampling point for the $k$-th sub-range in $l$-th layer.

For the selected sampling points, the next step is to find the corresponding codewords in the codebook. Therefore, we try to find a codeword ${\bf{g}}_f$ corresponding to the sampling point $\psi_{l,k}$ in ${\cal{G}}$. 
However, the storage space of the codebook is limited, and the situation where that none of the codeword's corresponding sampling point in the existing codebook is $\psi_{l,k}$ may happen.
In this case, generate a codeword ${\bf{g}}_{l,k}$ following steps S1 to S5,  and add it to the codebook ${\cal{G}}$.

Repeat the above process until $l=L_1$, and obtain the new sampling range along $x$-axis dimension, denoted by ${\cal{D}} = \left[x_{\rm{min}}^{\rm{opt}}, x_{\rm{max}}^{\rm{opt}}\right]$, which will be used for the second-stage training.

\begin{myRemark}
	{\textit{It is noteworthy that during the beam training process, the user needs to provide feedback on the training results to the BS, which then adjusts the beam direction based on this feedback. In other words, there is an uplink transmission involved. 
	In the PASS considered in this paper, we assume that the uplink feedback signals to the BS are transmitted through the wireless channels.}}
\end{myRemark}

\subsubsection{Stage 2 - Fine-grained estimation}
Through the training at the first stage, the user's location along the $x$-axis dimension is roughly determined. 
At the second stage, the hierarchical beam training scheme is employed, and the number of activated pinching antennas increases with the number of training layers, i.e., $N’ = \min \left\{ 2^ {l+1-L_1}, N \right\} $, where the layer index $l$ continues along the first-stage training. As the training layer increases, the transmitted beam gradually becomes more refined, resulting in higher training resolution.

Let $L_2$ be the number of training layers at the second stage. In each layer, the sampling range is divided into $K$ equal parts along the $y$ direction.
Since we focus on the $y$ direction at the second stage, we denote the sampling range by ${\cal{D}} = \left[y_{\rm{min}}, y_{\rm{max}}\right]$, and denote the $k$-th sub-range in $l$-th layer by ${\cal{D}}_{l,k} = \left[y_{\rm{min}}^{l,k}, y_{\rm{max}}^{l,k}\right]${\footnote{Similarly, the one-dimensional representation here is used for simplifying notation. The sampling area is a 2D plane, and the range in the $x$-axis dimension is from the beam training at the first stage, i.e., $\left[x_{\rm{min}}^{\rm{opt}}, x_{\rm{max}}^{\rm{opt}}\right]$.}}.
The midpoint of ${\cal{D}}_{l,k}$ is denoted by $\psi_{l,k} = \left(\frac{x_{\rm{min}}^{\rm{opt}}+x_{\rm{max}}^{\rm{opt}}}{2}, y_{l,k}, 0\right)$ with $y_{l,k} = \left(y_{\rm{min}}^{l,k}+y_{\rm{max}}^{l,k}\right)/2$.
The optimal sub-range ${\cal{D}}_{l,k}^{\rm{opt}}$ for each layer will serve as the sampling range for the next layer. 
That is, in the $\left(l+1\right)$-th layer, the sampling range ${\cal{D}}$ will be updated by ${\cal{D}}_{l,k}^{\rm{opt}}$.

The initial sampling range is ${\cal{D}} = \left[0,  D_y\right]$. 
In the $l$-th layer, ${\cal{D}}$ is divided into $K$ equal parts with length $\frac{D_y}{K^{l-L_1}}$ along the $y$ direction. 
The $k$-th sub-range in $l$-th layer is
\begin{equation}
	\begin{array}{l}
		{\cal{D}}_{l,k}  =\left[y_{\rm{min}}^{l,k},  y_{\rm{max}}^{l,k} \right] 
		=  \left[y_{\rm{min}}^{l-1,k} + \frac{\left(k-1\right)D_y}{K^{l-L_1}},  y_{\rm{min}}^{l-1,k} + \frac{kD_y}{K^{l-L_1}}\right],
	\end{array}
\end{equation}
where the midpoint is
\begin{equation}
	\begin{array}{l}
	\psi_{l,k} = \left(\frac{x_{\rm{min}}^{\rm{opt}}+x_{\rm{max}}^{\rm{opt}}}{2}, y_{\rm{min}}^{l-1,k} + \frac{kD_y}{K^{l-L_1}} - \frac{D_y}{2K^{l-L_1}}, 0\right),
	\end{array}
\end{equation}
which is the sampling point for the $k$-th sub-range in $l$-th layer.

Next, try to find a codeword ${\bf{g}}_f$ in ${\cal{G}}$, where ${\bf{g}}_f$'s corresponding sampling point ${\psi_f}$ is $\psi_{l,k}$.
Otherwise, generate a codeword ${\bf{g}}_{l,k}$ following steps S1 to S5,  and add it to the codebook ${\cal{G}}$.
Repeat the above process until $l=L_1 + L_2$, and obtain the sampling range ${\cal{D}} = \left[y_{\rm{min}}^{\rm{opt}}, y_{\rm{max}}^{\rm{opt}}\right]$.
At the end of the second stage, the sampling range for the third stage beam training can be obtained, i.e., ${\cal{D}} = \left\{\left[x_{\rm{min}}^{\rm{opt}}, x_{\rm{max}}^{\rm{opt}}\right],  \left[y_{\rm{min}}^{\rm{opt}}, y_{\rm{max}}^{\rm{opt}}\right]\right\}$.

\begin{algorithm}[!t]
	\caption{3SBT scheme for SWSU-PASS}
	\label{3SBT}
	\begin{algorithmic}[1]
		\begin{small}
		\REQUIRE number of activated pinching antennas $N$, sampling range $\cal{D}$, number of sampling points in each layer for hierarchical training $K$, training layers for first and second stages $L_1$ and $L_2$, threshold of sub-range length for exhaustive search $d_{\rm{ES}}$, existing codebook ${\cal{G}}$;
		\ENSURE optimal codeword ${\bf{g}}_{\rm{opt}} $, updated codebook ${\cal{G}}$;
		\STATE {\bf{Initialization:}} $l=1$, $\left|r\right|_{\rm{opt}} =0$;
		\STATE {\bf{Stage 1: Coarse-grained hierarchical estimation}}
		\WHILE{$l \le L_1$}
		\STATE  One pinching antenna is activated;
		\STATE Divide the sampling range ${\cal{D}}$ along $x$-axis dimension into $K$ equal parts, which are denoted by ${\cal{D}}_{l,k}$, $k\in \left\{1, \cdots, K\right\}$. The midpoint of ${\cal{D}}_{l,k}$ is denoted by $\psi_{l,k} = \left(x_{l,k}, 0, 0\right)$;
		\FORALL {$k\in \left\{1, \cdots, K\right\}$} \label{CoarseBegin}
		\STATE Find the codeword ${\bf{g}}_f$ in ${\cal{G}}$, where ${\bf{g}}_f$'s corresponding sampling point ${\psi_f}$ is $\psi_{l,k}$;
		\IF{${\bf{g}}_f$ cannot be found} 
		\STATE Generate a codeword ${\bf{g}}_{l,k}$ by following steps S{\ref{S1}} to S{\ref{S5}},  and add it to the codebook ${\cal{G}}$;
		\ENDIF
		\IF{$\left|r_{l,k}\right| > \left|r\right|_{\rm{opt}} $}
		\STATE $ \left|r\right|_{\rm{opt}} = \left|r_{l,k}\right| $, ${\bf{g}}_{\rm{opt}} = {\bf{g}}_{l,k} $;
		\STATE $x_{\rm{opt}} = x_{l,k}$; \label{x_opt}
		\STATE ${\cal{D}} \gets {\cal{D}}_{l,k}$;
		\ENDIF
    	\ENDFOR \label{CoarseEnd}
		\STATE  $l \gets l+1$;
		\ENDWHILE
		\STATE {\bf{Stage 2: Fine-grained hierarchical estimation}} 
		\WHILE{$l \le L_1 + L_2$}
		\STATE $N’= \min \left\{ 2^ {l+1-L_1}, N \right\} $ pinching antennas are activated;
		\STATE Divide the sampling range ${\cal{D}}$ along $y$-axis into $K$ equal parts, which are denoted by ${\cal{D}}_{l,k}$, $k\in \left\{1, \cdots, K\right\}$. The midpoint of ${\cal{D}}_{l,k}$ is denoted by $\psi_{l,k} = \left(x_{\rm{opt}}, y_{l,k}, 0\right)$, $k\in \left\{1, \cdots, K\right\}$;
		\STATE Repeat steps \ref{CoarseBegin}--\ref{CoarseEnd}, where step \ref{x_opt} is replaced by $y_{\rm{opt}} = y_{l,k}$;
		\STATE  $l \gets l+1$;
		\ENDWHILE
		\STATE {\bf{Stage 3: Partial exhaustive search}}
		\STATE $N$ pinching antennas are activated;
		\STATE Divide the sampling range $\cal{D}$ into $K_1 \times K_2$ equal sub-ranges, where each sub-range is denoted by ${\cal{D}}_{k_1, k_2}$, $k_1 \in \left\{1, \cdots, K_1\right\}$ and $ k_2 \in \left\{1, \cdots, K_2\right\}$. The midpoint of ${\cal{D}}_{k_1, k_2}$ is denoted by $\psi_{k_1,k_2} = \left(x_{k_1}, y_{k_2}, 0\right)$;
		\FORALL {$k_1\in \left\{1, \cdots, K_1 \right\}$ and $k_2\in \left\{1, \cdots, K_2 \right\}$} \label{ES_Begin}
		\STATE Find the codeword ${\bf{g}}_f$ in ${\cal{G}}$, where ${\bf{g}}_f$'s corresponding sampling point is $\psi_{k_1, k_2}$;
		\IF{${\bf{g}}_f$ cannot be found} 
		\STATE Generate a codeword ${\bf{g}}_{k_1, k_2}$ by following steps S{\ref{S1}} to S{\ref{S5}},  and add it to the codebook ${\cal{G}}$;
		\ENDIF
		\IF{$\left|r_{k_1,k_2}\right| > \left|r\right|_{\rm{opt}} $}
		\STATE $ \left|r\right|_{\rm{opt}} = \left|r_{k_1,k_2}\right| $, ${\bf{g}}_{\rm{opt}} = {\bf{g}}_{k_1,k_2} $;
		\STATE $x_{\rm{opt}} = x_{k_1}$, $y_{\rm{opt}} = y_{k_2}$;
		\STATE ${\cal{D}} \gets {\cal{D}}_{k_1,k_2}$;
		\ENDIF
		\ENDFOR \label{ES_End}
	\end{small}
	\end{algorithmic}
\end{algorithm}

\subsubsection{Stage 3 - Partial exhaustive search}  \label{Stage3_ES} 
An estimated user location has been obtained through the beam training in the first and second stages. 
However, limited spatial quantization makes estimation errors inevitable. 
Therefore, in the partial exhaustive search stage, we aim to achieve further beam alignment by employing an exhaustive approach.

Divide the sampling range obtained from the second stage into ${K}_1$ and ${K}_2$ equal parts along the $x$ and $y$ directions respectively, ensuring that the length of each sub-range is less than $d_{\rm{ES}}$.
The following constraints should be satisfied
\begin{equation}
	\begin{array}{l}
	{K_1} \ge \left \lceil \frac{x_{\rm{max}}^{\rm{opt}} - x_{\rm{min}}^{\rm{opt}}}{d_{\rm{ES}}} \right \rceil, \;\;
	{K_2} \ge \left \lceil \frac{y_{\rm{max}}^{\rm{opt}} - y_{\rm{min}}^{\rm{opt}}}{d_{\rm{ES}}} \right \rceil ,
	\end{array}
\end{equation}
where $\left \lceil \cdot \right \rceil$ represents the ceiling function.
Therefore, we obtain $K_1 \times K_2$ sampling sub-ranges, each with an area of $ S_{\rm{ES}} = \frac{D_x}{K^{L_1} {K_1}} \times \frac{D_y}{K^{L_2} {K_2}}$.

Denote the $\left \langle {k_1, k_2} \right \rangle $-th sampling sub-range by ${\cal{D}}_{k_1, k_2}$, where $k_1 \in \left\{1, \cdots, K_1\right\}$ and $ k_2 \in \left\{1, \cdots, K_2\right\}$ are the indexes for $x$ and $y$, respectively.
The ranges in the $x$ and $y$ directions of ${\cal{D}}_{k_1, k_2}$ are 
\begin{subequations}
\begin{equation}
	\begin{array}{l}
	\left[x_{\rm{min}}^{\rm{opt}} + \frac{\left(k_1 - 1\right)D_x}{K^{L_1} K_1}, x_{\rm{min}}^{\rm{opt}} + \frac{k_1 D_x}{K^{L_1} K_1}, \right]
	\end{array}
\end{equation}
and
\begin{equation}
	\begin{array}{l}
	\left[y_{\rm{min}}^{\rm{opt}} + \frac{\left(k_2 - 1\right)D_y}{K^{L_2} K_2}, y_{\rm{min}}^{\rm{opt}} + \frac{k_2 D_y}{K^{L_2} K_2}, \right]
	\end{array}
\end{equation}
\end{subequations}
respectively. 
The midpoint of ${\cal{D}}_{k_1, k_2}$ is denoted by 
\begin{equation}
\begin{array}{l}
	\psi_{k_1, k_2} =  \left(  x_{\rm{min}}^{\rm{opt}} + \frac{\left(k_1 - 1/2\right) D_x}{K^{L_1} K_1}, 
	 y_{\rm{min}}^{\rm{opt}} + \frac{\left(k_2- 1/2\right) D_y}{K^{L_2} K_2}, 0\right),
\end{array}
\end{equation}
which is the sampling point for the $\left \langle {k_1, k_2} \right \rangle $-th sub-range. 
For each sampling point, try to find a codeword ${\bf{g}}_f$ in ${\cal{G}}$, where ${\bf{g}}_f$'s corresponding sampling point is $\psi_{k_1, k_2}$.
Otherwise, generate a codeword ${\bf{g}}_{k_1, k_2}$ by following steps S1 to S5,  and add it to the codebook ${\cal{G}}$. 

After exhaustively searching through all $K_1 \times K_2$ sampling points, the optimal codeword can be obtained, which can be regarded as the searched globally optimal codeword in the three-stage beam training. 
The proposed 3SBT scheme for SWSU-PASS is summarized in {\bf{Algorithm \ref{3SBT}}}.

\section{Single-Waveguide PASS Serving Multiple Users} \label{1WG_mU}

In this section, we focus on the scenario of SWMU-PASS.
Firstly, we introduce the system model of NOMA-based SWMU-PASS. 
Subsequently, based on the scalable codebook generation design, we propose an improved 3SBT scheme. 
The details are presented in the following subsections, respectively.

\subsection{System Model of SWMU-PASS}

\begin{figure}[t]
	\centering
	\includegraphics[width=6.5cm]{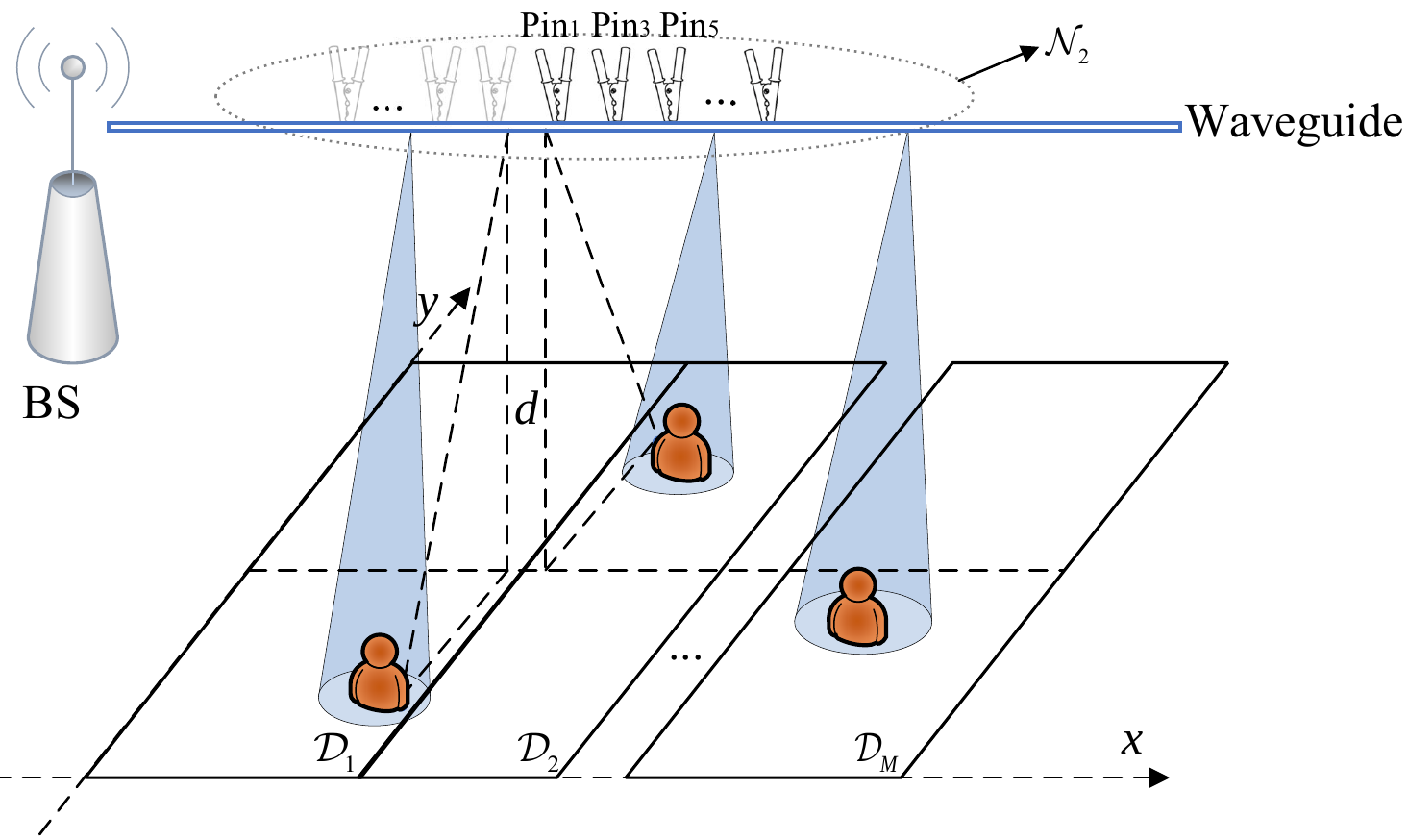}
	\caption{Illustration of SWMU-PASS.}
	\label{SystemModel_SWMU}
\end{figure}

Given that there is only one waveguide, only one data stream can be supported at one time. 
Therefore, a natural approach is to use orthogonal multiple access, such as time-division multiple access (TDMA), to serve multiple users, e.g., a single user is served in a single time slot. 
In this case, it is only necessary to execute the proposed single-user beam training scheme for each user in separated time slots.
However, to enhance spectral efficiency, we consider a multi-user beam training scheme that supports NOMA-assisted transmission.
Particularly, users that are far away from each other are scheduled for the implementation of NOMA, which also justifies the choice of $\tilde \psi _m^{\rm{pin}} = \psi _m^{\rm{pin}}$.

The $N$ pinching antennas can be activated on the single waveguide.
The number of users is $M$, where the $m$-th user is denoted by ${\rm{U}}_m$,  $1 \le m \le M$. 
It is assumed that ${\rm{U}}_m$ is located within a 2D plane ${\cal{D}}_m$.
Denote the location of ${\rm{U}}_m$ by $\psi _m = \left(x_m, y_m, 0\right)$, and the location of pinching antenna on the waveguide that is closest to ${\rm{U}}_m$ is denoted by $\psi _m^{\rm{pin}} = \left(x_m, 0, d\right)$.
To simplify the analysis, we assume that the regions from ${\cal{D}}_1$ to ${\cal{D}}_M$ are arranged along the $x$-axis dimension, as shown in Fig. {\ref{SystemModel_SWMU}}.

The transmitted superimposed signal is $s = \sum_{m=1}^{M} \sqrt{\alpha_m} s_m$,
where $s_m$ is the signal for ${\rm{U}}_m$, $\alpha_m$ is the power allocation coefficient for ${\rm{U}}_m$, and $\sum_{m=1}^{M} \alpha_m = 1$.
By taking the phase shifts caused by in-waveguide propagation into consideration, the channel vector of ${\rm{U}}_m$ located at $\psi_m = \left(x_m, y_m, 0\right)$ can be given as
\begin{equation}
	\begin{array}{l}
	{\bf{h}}_m = {\left[ {\frac{{\sqrt \eta  {e^{ - j\frac{{2\pi }}{\lambda }\left(\left| {{\psi}_m - \tilde \psi _1^{{\rm{pin}}}} \right| + \theta_1 \right)}}}}{{\left| {{\psi}_m - \tilde \psi _1^{{\rm{pin}}}} \right|}} \cdots \frac{{\sqrt \eta  {e^{ - j\frac{{2\pi }}{\lambda }\left(\left| {{\psi}_m - \tilde \psi _N^{{\rm{pin}}}} \right|+\theta_N \right)}}}}{{\left| {{\psi}_m - \tilde \psi _N^{{\rm{pin}}}} \right|}}} \right]^T}.
	\end{array}
\end{equation}
Similarly, consider the total power is equally divided among activated antennas, i.e., ${\bf{g}} = \left[\sqrt{\frac{P}{N}}, \cdots, \sqrt{\frac{P}{N}} \right] \in {\mathbb{C}}^{N \times 1}$.
Therefore, the received signal at ${\rm{U}}_m$ is
\begin{equation}\label{ReceivedSignal}
	v_m = {\bf{h}}_m^H {\bf{g}} \sum_{i=1}^{M} \sqrt{\alpha_i} s_i = {\bf{h}}_m^H {\bf{g}}  \sqrt{\alpha_m} s_m + {\bf{h}}_m^H {\bf{g}} \sum_{i\ne m} \sqrt{\alpha_i} s_i +  w_m,
\end{equation}
where $w_m \sim {\cal{CN}} \left(0, \sigma_m^2 \right)$ is the AWGN at ${\rm{U}}_m$. 
Eq. (\ref{ReceivedSignal}) can be viewed as a multi-user downlink NOMA system, where successive interference cancellation (SIC) can be utilized to remove multiple-access interference. The SIC decoding order can be established based on the effective channel gains.
Without loss of generality, we consider the channel conditions are ordered as follows
\begin{equation}
	\left|{\bf{h}}_1\right| \le \left|{\bf{h}}_2\right| \le \cdots \le \left|{\bf{h}}_M\right|.
\end{equation}
This channel order can be realized via an appropriate antenna activation approach, which will be discussed later.

According to the principle of downlink power-domain NOMA, ${\rm{U}}_m$ will first decode and remove ${\rm{U}}_i$'s signal, $1 \le i \le m-1$, and then decode its own signal by treating ${\rm{U}}_j$'s signal as interference, $j > m$.
Therefore, the received signal-to-interference-plus-noise ratio (SINR) of ${\rm{U}}_m$ is 
\begin{equation}
	\begin{array}{l}
	\gamma_m = \frac{\left|{\bf{h}}_m^H {\bf{g}}\right|^2\alpha_m}{  \left|{\bf{h}}_m^H {\bf{g}}\right|^2 \sum_{j=m+1}^{M}\alpha_j +\sigma_m^2}.
	\end{array}
\end{equation}
The achievable data rate of ${\rm{U}}_m$'s signal can be given by
\begin{equation}
	R_m = \min \left\{ R_{m,m}, R_{m+1,m}, \cdots, R_{M,m}	\right\},
\end{equation}
where $R_{j,m}$ denotes the data rate for ${\rm{U}}_j$ to decode ${\rm{U}}_m$'s signal, i.e., \cite{LSP.2014.2343971}
\begin{equation}
	\begin{array}{l}
	R_{j,m} = \log_2 \left(1+\frac{{\left|{\bf{h}}_j^H {\bf{g}}\right|^2\alpha_m}}{ \left|{\bf{h}}_j^H {\bf{g}}\right|^2\sum_{i=m+1}^{M}  \alpha_i +\sigma_j^2}\right), j \ge m.
	\end{array}
\end{equation}

\subsection{Improved 3SBT scheme}\label{Beamtraining2}

Note that the scalable codebook generation scheme proposed in {\bf{Algorithm \ref{alg:1}}} is still applicable for multi-user beam training in single-waveguide PASS. 

\subsubsection{Stage 1 - Separated user training} 

At the first stage of multi-user beam training, we consider separated training among users to roughly locate the positions of each user, thereby focusing on the desired signal strength of each user, given by
\begin{equation}
	\left|r_m\right| ={\left|{\bf{h}}_m^H {\bf{g}}\right|^2\alpha_m}.
\end{equation} 

We note that the free-space large-scale path loss term, $\left| {{\psi_m} - \tilde \psi _n^{{\rm{pin}}}} \right|$, is dominant compared to the other terms in $h_{m,n} = \frac{{\sqrt \eta  {e^{ - j\frac{{2\pi }}{\lambda }\left(\left| {{\psi}_m - \tilde \psi _n^{{\rm{pin}}}} \right| + \theta_n \right)}}}}{{\left| {{\psi}_m - \tilde \psi _n^{{\rm{pin}}}} \right|}}$. 
In order to maximize the desired signal strength, it is ideal if each user has activated pinching antennas located as close to them as possible. 
Therefore, it is natural to consider to divide all activated antennas into multiple clusters, and assign them for the training of each user separately, where details are given as follows. 

Divide the activated pinching antennas into $M$ clusters, where ${\cal{N}}_m$ $\left(\left|{\cal{N}}_m\right| = N_m\right)$ is the cluster of pinching antennas that are configured for ${\rm{U}}_m$, $m = 1, \cdots, M$.
Define $m' = \bmod \left\{ N, M\right\} $, then we have $N_m = \left \lceil \frac{N}{M} \right \rceil$ if $1\le m \le m'$, and $N_m = \left \lfloor \frac{N}{M} \right \rfloor$ if $m' < m \le M$, where $ \left \lceil \cdot \right \rceil$ and $\left \lfloor \cdot \right \rfloor$ represent the ceiling function and the floor function, respectively.
After antenna clustering, for each user ${\rm{U}}_m$ located at range ${\cal{D}}_m$, perform the hierarchical beam training for a single user, which follows the process in {\bf{Algorithm {\ref{3SBT}}}}.

We note that although different clusters of antennas are physically separated, they still perform beam training simultaneously. 
Each cluster of antennas conducts beam direction alignment within its corresponding sampling area. We note that a user associated to one cluster may also receive signals from other antenna clusters. However, due to the impact of large-scale fading, the signal strength from the user's antenna cluster still dominates. 
Furthermore, at this stage, we consider that the BS does not have specific markers or identifiers for each user during the beam training process. 
Therefore, there is no need to consider the impact caused by interference from other users' signals.
In fact, this stage can be seen as multiple individual users perform beam training  simultaneously within their respective sub-ranges, and the BS schedules the next training round based on the signal strength feedback from each user.
Through separated training, the updated sampling range of ${\rm{U}}_m$
\begin{equation}
	\begin{array}{l}
	{\cal{D}}_m =  \left\{\left[x_{m, \rm{min}}^{\rm{opt}}, x_{m,\rm{max}}^{\rm{opt}}\right],  \left[y_{m,\rm{min}}^{\rm{opt}}, y_{m,\rm{max}}^{\rm{opt}}\right]\right\}, 
	\end{array}
\end{equation} 
and roughly-estimated location of ${\rm{U}}_m$
\begin{equation}
	\psi_m^{\rm{opt}}= \left(x_m^{\rm{opt}}, y_m^{\rm{opt}}, 0\right), m = 1, \cdots, M
\end{equation}
can be obtained, which will be used in the second-stage multi-user beam training.

\subsubsection{Stage 2 - Partial antenna reclustering}
Considering a scenario where the distance between two adjacent users, ${\rm{U}}_{m-1}$ and ${\rm{U}}_{m}$, is extremely close, it is feasible to merge the two clusters of activated pinching antennas into one and perform beam training simultaneously for these two users.
The signals of these two users can be distinguished in the power domain{\footnote{By optimizing the power allocation strategy at the transmitter and designing the SIC order at the receiver, the users can distinguish and correctly decode their desired signals, and the overall performance of the system can also be optimized. However, the primary objective of the beam training phase is to determine the beam direction, therefore the optimization of power allocation and the design of SIC order are beyond consideration of this paper.}}.
That is, activated pinching antenna clusters ${\cal{N}}_{m-1}$ and ${\cal{N}}_m$ will be merged into one if $\left| \psi_{{m-1}}^{\rm{opt}} - \psi_{m}^{\rm{opt}} \right| \le \tilde d$ is satisfied, where $\tilde d$ is the distance threshold for reclustering.
Accordingly, the sampling ranges for ${\rm{U}}_{m-1}$ and ${\rm{U}}_{m}$ will be combined to form a single range.
The combined sampling range should cover ${\cal{D}}_{m-1}$ and ${\cal{D}}_m$, and should be a rectangular area for the subsequent  beam training with an exhaustive search. 
Therefore, the merged sampling range can be given by
\begin{equation}\label{NewRange}
	\begin{aligned}
		& {\cal{D}}_{\left[m-1, m\right]} =  \\
		& \bigg\{  \left[\min \left\{x_{m-1, \rm{min}}^{\rm{opt}}, x_{m, \rm{min}}^{\rm{opt}}\right\}, \max \left\{x_{m-1,\rm{max}}^{\rm{opt}}, x_{m,\rm{max}}^{\rm{opt}}\right\}\right], \\
		& \left[\min \left\{y_{m-1, \rm{min}}^{\rm{opt}}, y_{m, \rm{min}}^{\rm{opt}}\right\}, \max \left\{y_{m-1,\rm{max}}^{\rm{opt}}, y_{m,\rm{max}}^{\rm{opt}}\right\}\right]\bigg\}.  
	\end{aligned}
\end{equation} 

Considering another scenario, despite the fact that $\left| \psi_{{m-1}}^{\rm{opt}} - \psi_{m}^{\rm{opt}} \right| > \tilde d$, the distance between the two users ${\rm{U}}_{m-1}$ and ${\rm{U}}_{m}$ in the $x$-axis dimension is so close that their corresponding optimal antenna locations may overlap. 
In this case, we adjust the locations of several activated antennas within clusters ${\cal{N}}_{m-1}$ and ${\cal{N}}_m$ to to avoid antenna coupling. 
Because the antenna guard distance has already been considered during codebook design, even for the extreme case where the $x$-coordinates of ${\rm{U}}_{m-1}$ and ${\rm{U}}_{m}$ are the same, the condition $\left| \tilde \psi _{ 2_{m-1}}^{{\rm{pin}}} - \tilde \psi _{1_{m}}^{ {\rm{pin}}} \right| > \tilde \Delta $ still holds, where $\tilde \psi _{ 2_{m-1}}^{{\rm{pin}}}$ and $\tilde \psi _{1_{m}}^{ {\rm{pin}}}$ are the location of the second activated pinching antenna in ${\cal{N}}_{m-1}$ and the location of the first activated pinching antenna in ${\cal{N}}_{m}$, respectively.

The location of the largest odd-numbered antenna in antenna cluster ${\cal{N}}_{m-1}$ and the location of the largest even-numbered antenna in antenna cluster ${\cal{N}}_m$ are denoted by $\tilde \psi _{ N_{m-1}^{\rm{odd}}}^{{\rm{pin}}} $ and $\tilde \psi _{ N_{m}^{\rm{even}}}^{ {\rm{pin}}}$, respectively.
If $\left| \tilde \psi _{ N_{m-1}^{\rm{odd}}}^{{\rm{pin}}} - \tilde \psi _{ N_{m}^{\rm{even}}}^{ {\rm{pin}}} \right| < \tilde \Delta $, deactivate the odd-numbered antennas in ${\cal{N}}_{m-1}$ and activate the expanded even positions according to S\ref{S4}, and deactivate the even-numbered antennas in ${\cal{N}}_m$ and activate the expanded odd positions according to S\ref{S3}.
Take ${\cal{N}}_2$ as an example,  the antenna activation pattern is illustrated in Fig. {\ref{SystemModel_SWMU}}. 
Although the two clusters of antennas obtained in this way are close to each other, they can generate beams with different directions, allowing users to be distinguished in the spatial domain.

\subsubsection{Stage 3 - Joint multi-user training}

After the coarse user localization at the first stage and the antenna reclustering at the second stage, at the third stage, we employ an exhaustive search method to maximize the sum rate, i.e., $R_{\rm{sum}} = \sum_{m=1}^{M} R_m$, through joint training of multiple users.
For each sampling range obtained from the second stage, further division is conducted during the multi-user joint training stage to achieve better accuracy.
The range division approach is similar to that in Section \ref{Stage3_ES}, where ${\cal{D}}_m$ into $K_{m,1} \times K_{m,2}$ is divided into equal sub-ranges, with each sub-range denoted by ${\cal{D}}_{k_{m,1}, k_{m,2}}$, $k_1 \in \left\{1, \cdots, K_{m,1}\right\}$ and $ k_{m,2} \in \left\{1, \cdots, K_{m,2}\right\}$.  
By selecting one sub-area of each user, we can obtain a combination of sub-area samples for all users, denoted by 
$\left \langle {\cal{D}}_{k_{1,1}, k_{1,2}}, \cdots, {\cal{D}}_{k_{m,1}, k_{m,2}}, \cdots, {\cal{D}}_{k_{M,1}, k_{M,2}} \right \rangle $, $m = 1, \cdots ,M$ and $k_{m, 1} \in \left\{1, \cdots, K_{m,1}\right\}$ and $k_{m,2} \in \left\{1, \cdots, K_{m,2}\right\}$, which means there are $\prod_{m=1}^{M}\left(K_{m,1} \times K_{m,2}\right)$ sampling combinations{\footnote{Generally, to reduce complexity and enhance reliability, the number of users scheduled for NOMA is typically between 2 and 3, implying that the number of sampling combinations will not be unacceptably large.}}. 
In order to optimize the sum-rate, it is necessary to compare all sampling combinations, and the optimal sampling combination with the maximum sum rate can be found.

The proposed improved 3SBT scheme for SWMU-PASS is summarized in {\bf{Algorithm {\ref{Improved3SBT}}}}.

\begin{algorithm}[!t]
	\caption{Improved 3SBT scheme for SWMU-PASS}
	\label{Improved3SBT}
	\begin{algorithmic}[1]
		\begin{small}
		\REQUIRE number of activated pinching antennas $N$, number of users $M$, sampling range of each user ${\cal{D}}_m$, existing codebook ${\cal{G}}$, distance threshold $\tilde d$;
		\ENSURE optimal codewords ${\bf{g}}_m^{\rm{opt}}$, updated codebook ${\cal{G}}$;
		\STATE {\bf{Initialization:}} \label{Initialization}
		\STATE $\left|r_m\right|_{\rm{opt}} =0$, $\forall m = 1, \cdots, M$;
		\STATE Divide the available pinching antennas into $M$ clusters, where the number of $m$-th cluster pinching antennas is $N_m$, $m = 1, \cdots, M$;
		\IF {$1\le m \le \bmod \left\{ N, M\right\} $}
		\STATE $N_m = \left \lceil \frac{N}{M} \right \rceil$;
		\ELSE
		\STATE $N_m = \left \lfloor \frac{N}{M} \right \rfloor$;
		\ENDIF
		\STATE {\bf{Stage 1: Separate user training}}
		\FORALL {$m = 1, \cdots, M$}
		\STATE Perform the hierarchical training according to Stage 1 and Stage 2 in {\bf{Algorithm \ref{3SBT}}}, where the selection of antenna locations on the $m$-th cluster is based on the location of sampling points in ${\cal{D}}_m$;
		\STATE Update the estimated location of ${\rm{U}}_m$, denoted by $\psi_m^{\rm{opt}}= \left(x_m^{\rm{opt}}, y_m^{\rm{opt}}, 0\right)$;
		\STATE Update the sampling range of ${\rm{U}}_m$, denoted by ${\cal{D}}_m$;
		\ENDFOR
		\STATE {\bf{Stage 2: Partial antenna reclustering}} 
		\FORALL{$m = 2, \cdots, M$} \label{ReclusterBegin}
		\IF{$\left| \psi_{{m-1}}^{\rm{opt}} - \psi_{m}^{\rm{opt}} \right| \le \tilde d$}
		\STATE ${\cal{N}}_{\left[m-1, m\right]} \gets {\cal{N}}_m \cup {\cal{N}}_{m-1}$;
		\STATE Update ${\cal{D}}_{\left[m-1, m\right]}$ according to (\ref{NewRange});		
		\ELSIF{$\left| x_{{m-1}}^{\rm{opt}} - x_{m}^{\rm{opt}} \right| \le \tilde d$}
		\IF{$\left| \tilde \psi _{ N_{m-1}^{\rm{odd}}}^{{\rm{pin}}} - \tilde \psi _{ N_{m}^{\rm{even}}}^{ {\rm{pin}}} \right| < \tilde \Delta $}
		\STATE Deactivate the odd-numbered pinching antennas in ${\cal{N}}_{m-1}$ and activate the expanded even positions;
		\STATE Deactivate the even-numbered pinching antennas in ${\cal{N}}_m$ and activate the expanded odd positions;
		\ENDIF 
		\ENDIF
		\ENDFOR  \label{ReclusterEnd}
		\STATE {\bf{Stage 3: Multi-user joint training}}
		\FORALL{${\rm{U}}_m$, $m = 1, \cdots, M$} \label{JointTrainingBegin}
		\STATE Divide the sampling range of each user ${\cal{D}}_m$ into $K_{m,1} \times K_{m,2}$ equal sub-ranges, where each sub-range is denoted by ${\cal{D}}_{k_{m,1}, k_{m,2}}$, $k_{m, 1} \in \left\{1, \cdots, K_{m,1}\right\}$ and $ k_{m,2} \in \left\{1, \cdots, K_{m,2}\right\}$. The midpoint of ${\cal{D}}_{k_{m,1}, k_{m,2}}$ is denoted by $\psi_{k_{m,1},k_{m,2}} = \left(x_{k_{m,1}}, y_{k_{m,2}}, 0\right)$;
		\ENDFOR
		\FORALL{$\prod_{m=1}^{M}\left(K_{m,1} \times K_{m,2}\right)$ sampling sub-range combinations}
		\STATE For each combination of sampling sub-ranges $\left \langle {\cal{D}}_{k_{1,1}, k_{1,2}}, \cdots, {\cal{D}}_{k_{m,1}, k_{m,2}}, \cdots, {\cal{D}}_{k_{M,1}, k_{M,2}} \right \rangle $, perform exhaustive training according to Stage 3 in {\bf{Algorithm \ref{3SBT}}} in each sub-range ${\cal{D}}_{k_{m,1}, k_{m,2}}, m = 1, \cdots, M$;
		\STATE Calculate $R_{\rm{sum}}$ and find the maximum one, and update the corresponding codeword ${\bf{g}}_m^{\rm{opt}}$ for each antenna cluster;
		\ENDFOR \label{JointTrainingEnd}
	\end{small}
	\end{algorithmic}
\end{algorithm}

\section{Multi-Waveguide PASS Serving Multiple Users} \label{mWG_mU}

In this section, we focus on the scenario of MWMU-PASS. 
Firstly, we introduce the system model of MWMU-PASS, where a generalized  expression of the received signal is presented. 
Secondly, we design an increased-dimensional scalable codebook.
Subsequently, we propose an increased-dimensional 3SBT scheme. 
The details are presented in the following subsections, respectively.

\subsection{System Model of MWMU-PASS}

\begin{figure}[t]
	\centering
	\includegraphics[width=7.0cm]{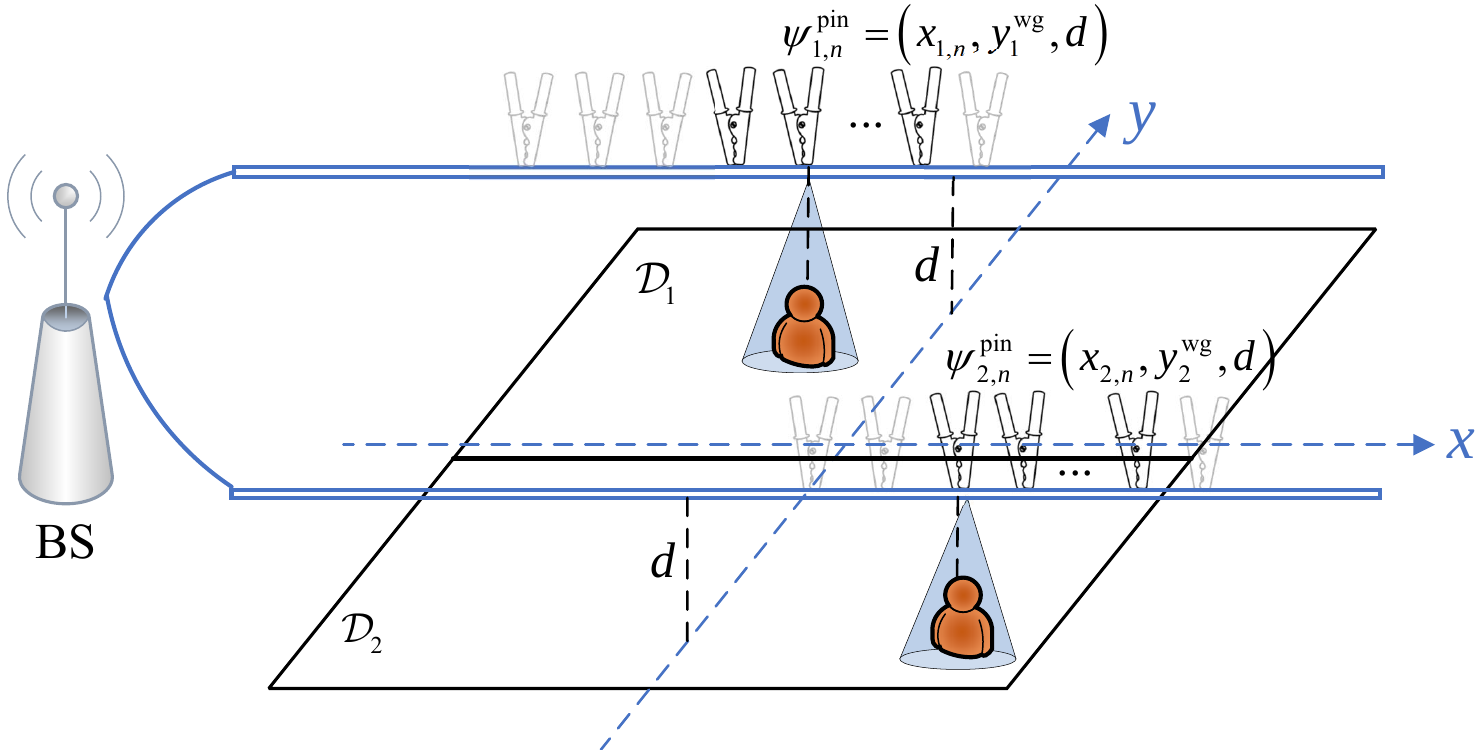}
	\caption{Illustration of MWMU-PASS.}
	\label{SystemModel_MWMU}
\end{figure}

Consider the number of waveguides is $Q$, and $N$ pinching antennas are activated on each waveguide, as shown in Fig. {\ref{SystemModel_MWMU}}. 
All the waveguides are placed parallel to the $x$-axis with a height of $d$, where the $y$-coordinate of the $q$-th waveguide is $y_q^{\rm{wg}}$.
Denote the $n$-th pinching antenna on the $q$-th waveguide by $\left\langle{q,n} \right\rangle$-th antenna, with its location denoted by ${\tilde \psi_{q,n}^{\rm{pin}}} $, where $q = 1, \cdots, Q$ and $n = 1, \cdots, N$. 

The number of users is $M$. 
We assume that the $m$-user is located within a 2D region ${\cal{D}}_m$, and the regions from ${\cal{D}}_1$ to ${\cal{D}}_M$ are arranged along the $y$-axis{\footnote{The arrangement of the sampling areas, whether along the $x$-axis dimension or the $y$-axis, is actually an artificial division for the convenience of analysis, and therefore does not lose generality.}}. 
Generally, each waveguide at the BS can support an independent data stream. The number of users simultaneously served by the BS is smaller than the number of waveguides, i.e., $M \le Q$. 
The considered scenario is similar to a commonly encountered situation in conventional-antenna systems, where the number of radio-frequency chains is no smaller than the number of served users. 
In this paper, we consider a case of $M = Q$. 
The channel between ${\rm{U}}_m$ and the $\left\langle {q,n} \right\rangle$-th antenna is
\begin{equation}
	\begin{array}{l}
	h_{m,q,n}=\frac{\sqrt{\eta}e^{-2\pi j \left( \frac{1}{\lambda }\left| \psi_m -\tilde\psi_{q,n}^{\rm{pin}}\right| + \frac{1}{\lambda_g }\left| \psi_0^{\rm{pin}} -\tilde\psi_{q,n}^{\rm{pin}}\right| \right)}}{\left | \psi_m -\tilde\psi_{q,n}^{\rm{pin}}\right |},
	\end{array}
\end{equation}
where the phase shifts $e^{-2\pi j \left( \frac{1}{\lambda }\left| \psi_m -\tilde\psi_{q,n}^{\rm{pin}}\right|  \right)}$ are caused by signals' propagation from the antennas to the users, and $e^{-2\pi j \left( \frac{1}{\lambda_g }\left| \psi_0^{\rm{pin}} -\tilde\psi_{q,n}^{\rm{pin}}\right| \right)}$ are caused by signals' propagation through the waveguide.
The channel vector from the $q$-th waveguide to ${\rm{U}}_m$ is 
\begin{equation}
	{\bf{h}}_{m, q} = \left[ h_{m,q,1}, \cdots, h_{m,q,n}, \cdots h_{m,q,N} \right].
\end{equation}

\begin{figure}[t]
	\centering
	\includegraphics[width=6.5cm]{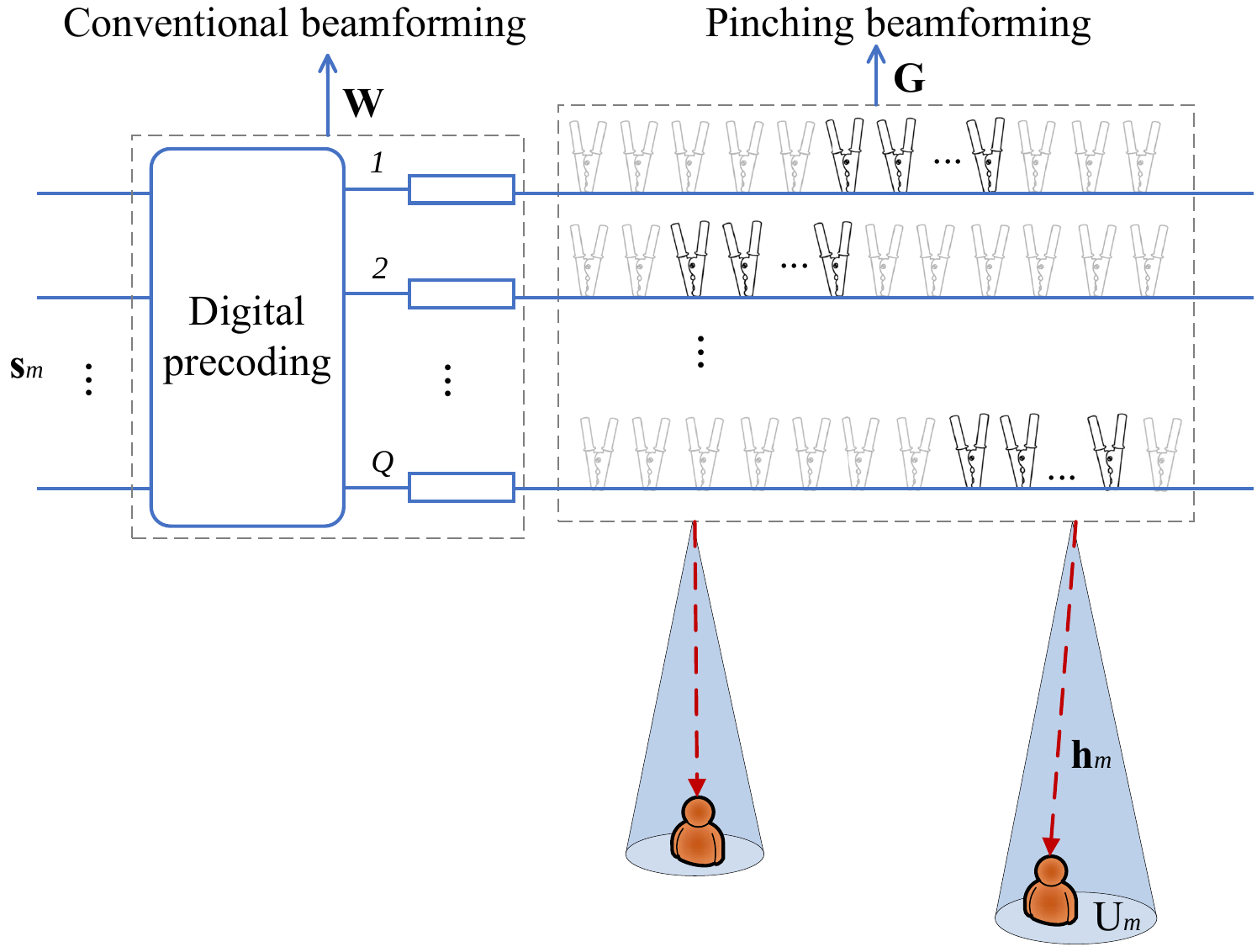}
	\caption{Illustration of partially-connected hybrid beamforming structure.}
	\label{HybridBeamforming}
\end{figure}

Different waveguides can be fed with different data streams, which means that partially-connected hybrid beamforming can be executed {\cite{TWC.2014.011714.130846, TCOMM.2022.3202215}}, as shown in Fig. {\ref{HybridBeamforming}}.  
Therefore, the received signal vector at ${\rm{U}}_m$ can be expressed as \begin{equation}\label{Signal_Um}
	\begin{array}{l}
	{\bf{v}}_m = {\bf{h}}_m^H  {\bf{G}} {\bf{W}}_m {\bf{s}}_m +\sum_{i\ne m} {\bf{h}}_m^H  {\bf{G}} {\bf{W}}_i {\bf{s}}_i +{\bf{w}}_m,
	\end{array}
\end{equation}
where ${\bf{h}}_m = \left[{\bf{h}}_{m, 1}, \cdots, {\bf{h}}_{m, q}, \cdots, {\bf{h}}_{m, Q}\right]^H \in {\mathbb{C}}^{QN \times 1}$ is the channel vector from $QN$ activated pinching antennas to ${\rm{U}}_m$, ${\bf{G}} \in {\mathbb{C}} ^{QN \times Q}$ is the pinching beamformer, ${\bf{W}}_m \in {\mathbb{C}}^{Q \times 1}$ is the digital precoder of the BS for ${\rm{U}}_m$, 
and ${\bf{s}}_m = \left[s_{m,1}, \cdots, s_{m,L_s}\right]^T \in {\mathbb{C}}^{L_s \times 1}$ is the signal vector to ${\rm{U}}_m$, with $L_s$ denoting the symbol length, satisfying $L_s \le Q$.  
${\bf{w}}_m \sim \left({\bf{0}}, \sigma_m^2 {\bf{I}} \right)\in {\mathbb{C}}^{L_s \times 1} $ denotes the AWGN at ${\rm{U}}_m$.
The SINR of ${\rm{U}}_m$ is given by
\begin{equation}\label{SINR_m}
	\begin{array}{l}
	\gamma_m = \frac{\left|{\bf{h}}_m^H  {\bf{G}} {\bf{W}}_m \right|^2}{ \left|\sum_{i\ne m} {\bf{h}}_m^H  {\bf{G}} {\bf{W}}_i\right|^2   +\sigma_m^2},
	\end{array}
\end{equation}
and the achievable sum rate can be given by
\begin{equation}\label{SumRate}
	\begin{array}{l}
	R_{\rm{sum}} = \sum_{m=1}^{M} \log_2 \left(1+ \gamma_m\right).
	\end{array}
\end{equation}

The codebook-based beam training avoids directly using ${\bf{h}}_m$, since the estimation of ${\bf{h}}_m$ incurs a large overhead, which means that ${\bf{h}}_m$ is not available during the beam training stage \cite{TWC.2020.3019523}. 
However, (\ref{SINR_m}) and (\ref{SumRate}) show that the received SINR at ${\rm{U}}_m$ and the achievable sum rate are closely related to the beamforming coefficients and channel conditions. 
In addition to its desired signals, ${\rm{U}}_m$ will also receive interference from signals of other users.
Furthermore, the strength and phase of the received signal at ${\rm{U}}_m$ are also strongly coupled with locations of the activated antennas, and they are  closely related to the channel conditions. 
It is challenging to obtain globally optimal performance through the optimization of antenna locations. 
Therefore, our focus is on acquiring CSI and performing beam alignment through efficient beam training schemes, without considering the optimization of digital precoder and pinching beamformer. 
To simplify the analysis, in this paper, we consider a scheme where total power on each waveguide is equally divided among the activated antennas. That is, if total transmitting power is $P$, and there are $QN$ activated pinching antennas, then the transmitting power of each antenna is $\frac{P}{QN}$. 
Based on the above considerations, we propose the following codebook generation design and beam training scheme for multi-waveguide PASS.

\subsection{Increased-Dimensional Scalable Codebook Generation} \label{Codebook2}

In addition to the degrees of freedom (DoFs) in signal transmission and beamforming, we note that the multi-waveguide PASS introduces additional DoFs in selecting the antenna locations compared to the single-waveguide PASS. 
Specifically, for a single-waveguide PASS, the activated antenna locations can only be chosen along the $x$-axis dimension at fixed $y$-coordinates. However, for a multi-waveguide PASS, the $y$-coordinates of the activated antenna locations can be changed by selecting different waveguides.
To begin with, we discuss the impact of antenna activation distribution on the received signal strength for a single user.

\begin{myLemma}\label{MultipleWaveguide_y}
	{\textit{Suppose that the user is located at $\psi_{\rm{U}} = \left(x_{\rm{U}}, y_{\rm{U}}, 0\right)$, the number of the activated pinching antennas is $N$, and the total power $P$ is equally shared among the activated antennas.
	If $\left| y_i^{\rm{wg}} - y_{\rm{U}}\right| < \left| y_j^{\rm{wg}} - y_{\rm{U}} \right|$, $\forall$ $i, j = 1, \cdots, Q$ and $i \ne j$, the user achieves the largest received signal strength when $N$ activated pinching antennas are all deployed on the $i$-th waveguide and cluster around $\psi^{\rm{pin}}_{i,{\rm{U}}} = \left(x_{\rm{U}}, y_i^{\rm{wg}}, d\right)$ by following S\ref{S1}--S\ref{S5}. }}
\end{myLemma}

{\textit{Proof:}} Please refer to Appendix {\ref{AppendixB}}. 
$\hfill\blacksquare$

Based on {\textit{Lemma {\ref{MultipleWaveguide_y}}}}, in order to maximize the desired received signal for users, it is ideal that each user has the antenna closest to them. 
Furthermore, in order to enhance the directionality of the beam, we consider clustering the pinching antennas on each waveguide together rather than separating them into small clusters.
However, the signal received by a user not only contains its own desired signal but also interference from other users' signals. 
In addition, {\textit{Lemma {\ref{MultipleWaveguide_y}}}} only considers the signal strength maximization of a single user.
Therefore, considering the overall performance and fairness among multiple users, we present the following codebook design for multi-user PASS.

Suppose there are $M$ sets of sampling points ${\cal{F}}_m \in {\cal{D}}_m$, $m \in \left\{1, \cdots, M\right\}$, a set of waveguides ${\cal{Q}}$, $q \in \left\{1, \cdots, Q\right\}$, with $M = Q$, and $N$ activated pinching antennas on each waveguide. 
Without loss of generality, we assume that $y_{1,\min}^{\cal{D}}>y_{2,\min}^{\cal{D}} >\cdots>y_M^{\cal{D}}$ and $y_1^{\rm{wg}} > y_2^{\rm{wg}}> \cdots >y_Q^{\rm{wg}}$ are satisfied, where $y_{m,\min}^{\cal{D}}$ is the minimum value of the $y$-direction coordinate in the $m$-th sampling range ${\cal{D}}_m$, $y_q^{\rm{wg}}$ is the $y$-coordinate of the $q$-th waveguide.
The multi-waveguide codebook can be generated through the following steps:
\begin{itemize}
	\item For each given sampling point located at ${\psi_{f,m}} = \left(x_{f,m}, y_{f,m}, 0\right) \in {\cal{D}}_m$, first identify its corresponding $m$-th waveguide.
	\item On the given $m$-th waveguide, find the locations of activated pinching antennas that are closest to the sampling point ${\psi_{f,m}}$ by following steps S{\ref{S1}} to S{\ref{S5}}.
	\item Repeat the above process until all the sampling points in ${\cal{F}}_m$ have generated their corresponding codewords, and each codeword has been stored in the codebook ${\cal{G}}_m$. 
	\item Repeat the above process until all sampling points in all sampling ranges have generated their corresponding codewords and are stored in the codebooks.
	\item As the numbers of the sampling points and the activated pinching antennas increases, the proposed multi-waveguide codebook is also scalable by following similar approaches in {\bf{Algorithm \ref{alg:1}}}, while only requiring an additional step to identify its corresponding waveguide.
\end{itemize}

\begin{myRemark}
	{\textit{Based on the proposed increased-dimensional codebook, the overall system performance, which can be measured by either the sum rate, user fairness, or other metrics, can be further optimized through an appropriate design of a digital precoder matrix. For example, if the beamforming coefficients satisfy $\alpha_{m,m} \ge \max \left\{\alpha_{i,m}\right\}, \forall i \ne m$, on the $m$-th waveguide, the signal power allocated to ${\rm{U}}_m$ is larger than the signal power allocated to other users, which means that the strength of  ${\rm{U}}_m$'s intended signal, $s_m$, is expected to be stronger than those of the interference signals.}}
\end{myRemark}

\subsection{Increased-Dimensional 3SBT Scheme} \label{Beamtraining3}

The main differences between the beam training schemes for multi-waveguide PASS and single-waveguide PASS are:
\begin{itemize}
 	\item Compared to the single-waveguide PASS case, in which the $y$-coordinates of pinching antennas are the same, each waveguide in the multi-waveguide PASS has a different $y$-coordinate. Therefore, when determining the locations of the activated pinching antennas, the impact of the large scale pathloss on the $y$-axis dimension needs to be comprehensively considered.
 	\item Different data streams can be passed through different waveguides, and hence the signal received by the user is a superposition of signals from the multiple waveguides, e.g., users receive their desired signal as well as interference signals from other users.
 	\item The optimization of overall system performance becomes more complicated, the key system factors, such as activated antenna locations, digital precoder, and power allocation, are highly coupled, which will affect the overall system performance.
\end{itemize}

In this subsection, our objective is to propose an effective beam training scheme that improves the accuracy and efficiency of beam alignment.
Therefore, we consider first performing separated beam training for each user to roughly determine their locations, and then executing multi-user joint beam training to optimize the sum-rate for all users. 
This process is similar to the case of SWMU-PASS, with the difference being that the locations of multiple waveguides need to be considered. 
By executing the proposed beam training scheme, the CSI of the users can be obtained, and the sum-rate is enhanced to a certain degree.
Nevertheless, the overall system performance can be further enhanced through alternating optimization with the beamformer design and power allocation, which is beyond the scope of this paper.
Based on these considerations, we propose an increased-dimensional 3SBT scheme for MWMU-PASS, as summarized in {\bf{Algorithm {\ref{Increased3SBT}}}}.

\begin{algorithm}[!t]
	\caption{Increased-dimensional 3SBT scheme for MWMU-PASS}
	\label{Increased3SBT}
	\begin{algorithmic}[1]
		\begin{small}
		\REQUIRE number of waveguides $Q$, number of activated pinching antennas $N$ on each waveguide, number of users $M$, sampling range for each user ${\cal{D}}_m$, existing codebook ${\cal{G}}_m$;
		\ENSURE optimal codewords ${\bf{g}}_m^{\rm{opt}}$, updated codebook ${\cal{G}}_m$;
		\STATE {\bf{Initialization:}}  \STATE $\left|r_m\right|_{\rm{opt}} =0$, $\forall m = 1, \cdots, M$;
		\STATE {\bf{Stage 1: Waveguide determination}}
		\STATE Sort the sampling areas ${\cal{D}}_m, m = 1, \cdots, M$ according to their coordinates in the $y$-direction in descending order, satisfying $y_{1,\min}^{\cal{D}}>y_{2,\min}^{\cal{D}} >\cdots>y_M^{\cal{D}}$;
		\STATE  Sort the waveguides according to their coordinates in the $y$-direction $y_q^{\rm{wg}}, q = 1, \cdots, Q$ in descending order, satisfying $y_1^{\rm{wg}} > y_2^{\rm{wg}}> \cdots >y_Q^{\rm{wg}}$;
		\FORALL {$m = 1, \cdots, M$}
		\STATE Associate the $m$-th waveguide with the $m$-th sampling range;
		\ENDFOR
		\STATE {\bf{Stage 2: Separated user training}}
		\FORALL {$m = 1, \cdots, M$}
		\STATE Perform the hierarchical training according to Stage 1 and Stage 2 in {\bf{Algorithm {\ref{3SBT}}}}, where the selection of antenna locations on the $m$-th waveguide is based on the location of sampling points in ${\cal{D}}_m$;
		\STATE Update the estimated location of ${\rm{U}}_m$, denoted by $\psi_m^{\rm{opt}}= \left(x_m^{\rm{opt}}, y_m^{\rm{opt}}, 0\right)$;
		\STATE Update the sampling range of ${\rm{U}}_m$, denoted by ${\cal{D}}_m$;
		\ENDFOR
		\STATE {\bf{Stage 3: Multi-user joint training}}
		\STATE Similar to steps {\ref{JointTrainingBegin}}--{\ref{JointTrainingEnd}} in {\bf{Algorithm {\ref{Improved3SBT}}}}.
	\end{small}
	\end{algorithmic}
\end{algorithm}

\section{Simulation Results}

In this section, we evaluate the performance of the pinching-antenna system through simulations. 
The parameter settings are as follows. 
The noise power is set as $-90$ dBm, the height of waveguide is set as $d = 3$ m, the carrier frequency is $f_c = 28$ GHz, the guard distance is $\tilde \Delta = \frac{\lambda}{2}$, and $n_{\rm{eff}} = 1.4$ \cite{Microwave}. 
The number of activated pinching antennas is 18.
The number of hierarchical training layers in the first and second stages are set as $L_1 = L_2 = 8$, the number of sampling points in each layer is $K=2$, and the length threshold of exhaustive searching sub-range is $d_{\rm{ES}} = 10^{-2}$.
The following three subsections will evaluate the cases of SWSU, SWMU, and MWMU, respectively. 
Table {\ref{Overhead}} compares the training overhead of various schemes to achieve a same training accuracy.

\begin{table*}[!t]
	\caption{Comparison of beam training overhead of various schemes}
	\begin{center}
		\begin{tabular}{|l|l|l|}
			\hline
			\rule{0pt}{9pt}
			{\bf{Beam training scheme}} & {\bf{Total training overhead}} & {\bf{In our set up}} \\
			\hline
			SWSU, 2D exhaustive search & $K^{L_1+L_2}\cdot{K_1}{K_2}$ & $2^{20}$  \\
			\hline
			SWSU, proposed & $K\left(L_1+L_2\right)+{K_1}{K_2}$ & 48  \\ 
			\hline
			SWMU, 2D exhaustive search & $K^{M\left(L_1+L_2\right)} \cdot \prod_{m=1}^{M}\left(K_{m,1} K_{m,2}\right)$ & $ 2^{40}, M=2; \;\;  2^{60}, M=3$  \\
			\hline
			SWMU, proposed & $MK\left(L_1+L_2\right)+\prod_{m=1}^{M}\left(K_{m,1} K_{m,2}\right)$ & $320, M=2; \;\; 4192, M=3$  \\ 
			\hline
			MWMU, 2D exhaustive search & $K^{M\left(L_1+L_2\right)} \cdot \prod_{m=1}^{M}\left(K_{m,1} K_{m,2}\right)$ & $2^{40}$  \\
			\hline
			MWMU, proposed & $MK\left(L_1+L_2\right)+\prod_{m=1}^{M}\left(K_{m,1} K_{m,2}\right)$ & 320  \\ 
			\hline
		\end{tabular}
		\label{Overhead}
	\end{center}
\end{table*}

\subsection{The SWSU case}

\begin{figure*}[t]
	\centering
	\begin{minipage}[t]{0.32\textwidth}
		\centering
		\includegraphics[width=6.0cm]{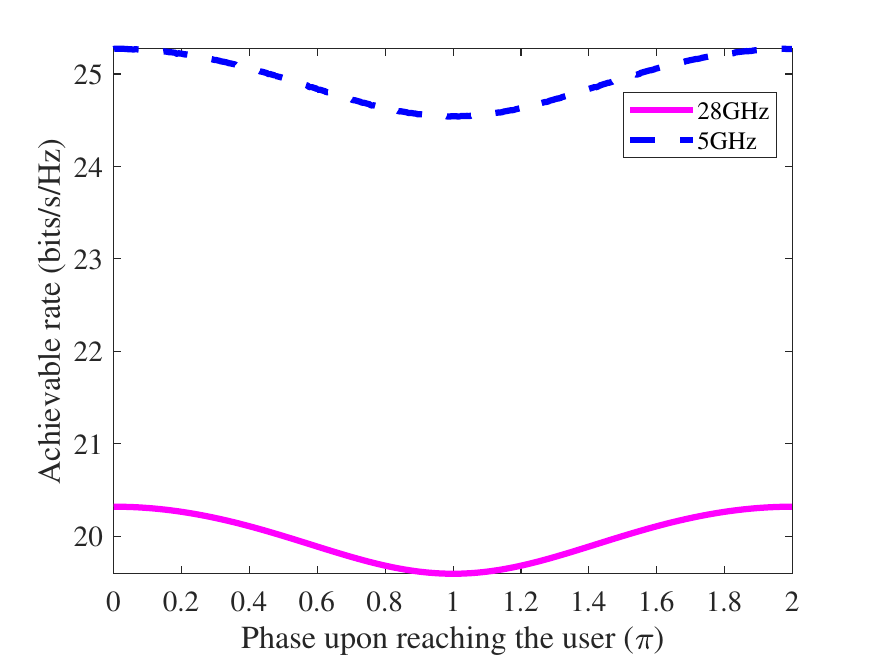}
		\caption{Achievable rate versus phase upon reaching the user, i.e., $\mod \left\{\frac{2\pi}{\lambda }\left| \psi_{\rm{U}} -\tilde\psi_{i,n}^{\rm{pin}}\right| +\theta_n , 2\pi\right\}$.}
		\label{Fig_1WG1U_Pattern}
	\end{minipage}
	\begin{minipage}[t]{0.32\textwidth}
		\centering
		\includegraphics[width=6.0cm]{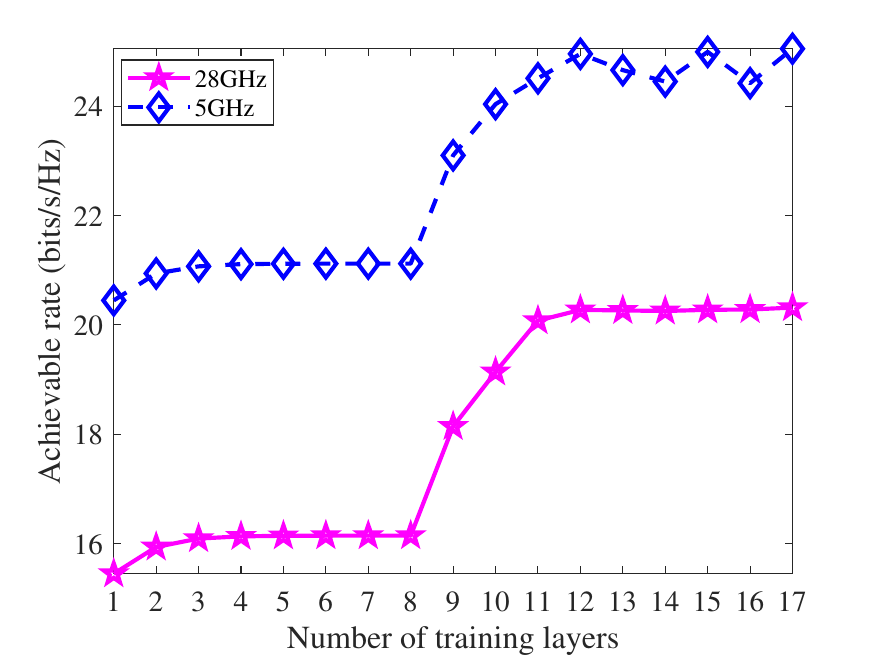}
		\caption{The variation of achievable rate with the number of training layers.}
		\label{Fig_1WG1U_TrainingLayer}
	\end{minipage}
	\begin{minipage}[t]{0.32\textwidth}
		\centering
		\includegraphics[width=6.0cm]{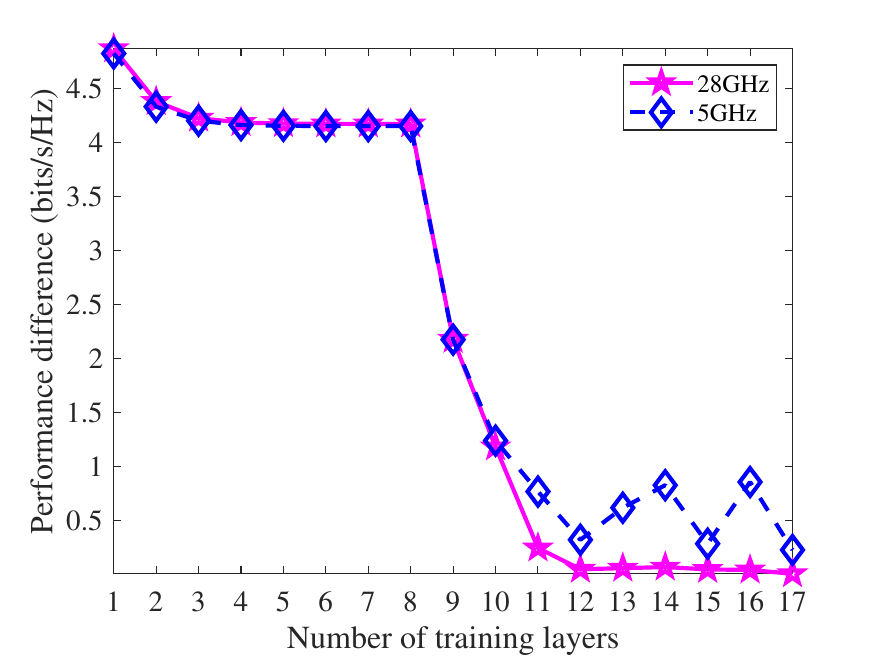}
		\caption{Performance difference between the obtained rate and the achievable rate under phase alignment versus the number of training layers.}
		\label{Fig_1WG1U_Difference}
	\end{minipage}
\end{figure*}

The user is located at ${\psi}_{\rm{U}} = \left(5, 4, 0\right)$, the sampling range is ${\cal{D}} =  \left\{\left[0, 10\right],  \left[0, 10\right]\right\}$, and the feed point of the waveguide is $\left(0, 5, 3\right)$, which is also the location of the BS. 
Fig. {\ref{Fig_1WG1U_Pattern}} illustrates the relationship between the achievable rate and the phase upon reaching the user. 
It can be observed that the achievable rate is maximized when the phase of the signals from all antennas reaching the user is 0, i.e., $\mod \left\{\frac{2\pi}{\lambda }\left| \psi_{\rm{U}} -\tilde\psi_{i,n}^{\rm{pin}}\right| + \theta_n , 2\pi\right\} = 0$. The rate decreases as the phase changes from 0 to $\pi$, and then increases as the phase changes from $\pi$ to 2$\pi$, returning to the performance at a phase of 0.
Additionally, the achievable rate at a carrier frequency of 5GHz is greater than of 28GHz. This is because $\eta$ is inversely proportional to the square of the frequency, and a higher frequency results in a larger channel attenuation.

Fig. {\ref{Fig_1WG1U_TrainingLayer}} and Fig. {\ref{Fig_1WG1U_Difference}} illustrate the training performance versus the number of beam training layers. 
Note that the exhaustive search training at the third stage is denoted by the $\left(L_1+L_2+1\right)$-th layer. 
In the first training stage, only one pinching antenna is activated.
Therefore, as shown in Fig. {\ref{Fig_1WG1U_TrainingLayer}}, the achievable data rate is relatively low, and the increase in the data rate is not significant from the fourth layer. At this stage, only a coarse localization of the user's position along the $x$-axis dimension can be achieved.
In the second training stage, starting from $l=9$, there is a noticeable increase in the achievable rate. This is due to the following reasons. On the one hand, the number of activated antennas increases with the number of layers. On the other hand, better beam alignment is achieved.
Fig. {\ref{Fig_1WG1U_Difference}} shows the performance difference between the obtained rate and the achievable rate under phase alignment, i.e., phase 0 in Fig. {\ref{Fig_1WG1U_Pattern}}. 
It can be observed that in the first few layers of beam training, the difference between the frequencies of 28 GHz and 5 GHz is not obvious. 
However, in the final layers of beam training, if the frequency is 28 GHz, the gap between the obtained data rate and the achievable data rate at phase alignment gradually narrows to nearly zero. In contrast, when the frequency is 5 GHz, the gap remains relatively significant and does not show a trend of converging.
This observation is due to the fact that longer wavelengths correspond to larger guard distances and discretization errors, resulting in increased phase alignment errors.

\begin{figure*}[t]
	\centering
	\begin{minipage}[t]{0.32\textwidth}
		\centering
		\includegraphics[width=6.0cm]{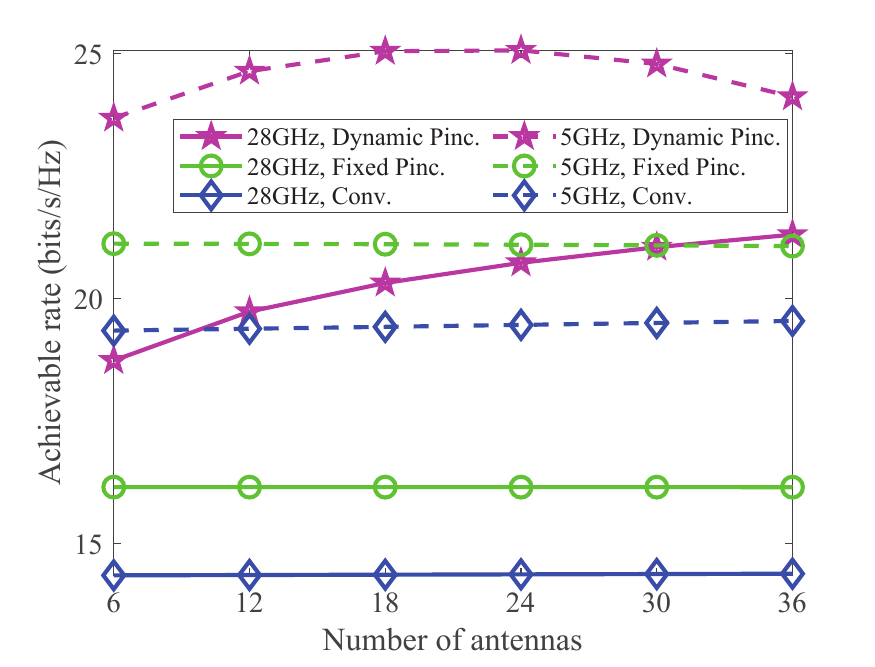}
		\caption{Achievable rate versus number of antennas.}
		\label{Fig_1WG1U_NumAntenna}
	\end{minipage}
	\begin{minipage}[t]{0.32\textwidth}
		\centering
		\includegraphics[width=6.0cm]{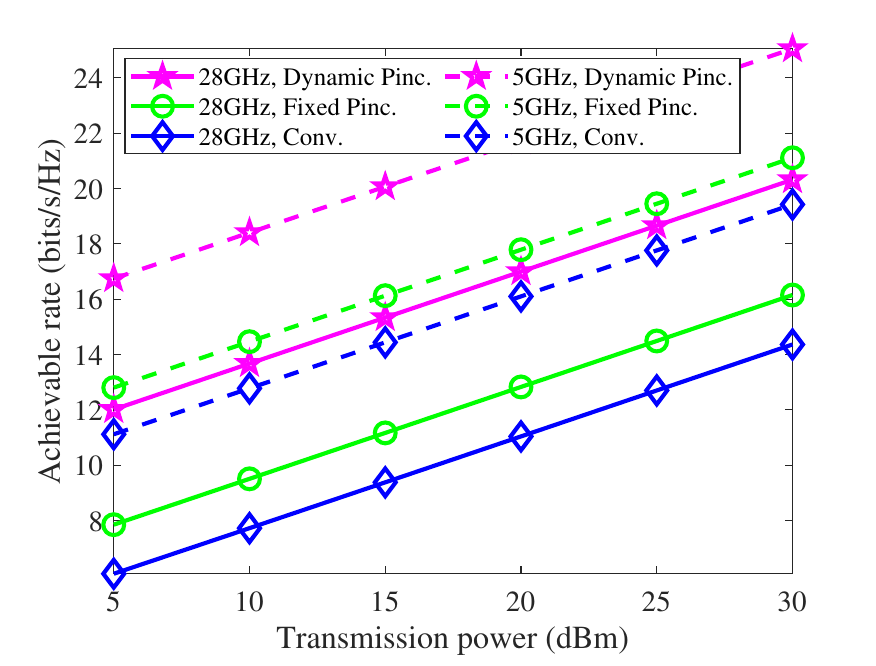}
		\caption{Achievable rate versus transmission power.}
		\label{Fig_1WG1U_Power}
	\end{minipage}
	\begin{minipage}[t]{0.32\textwidth}
		\centering
		\includegraphics[width=6.0cm]{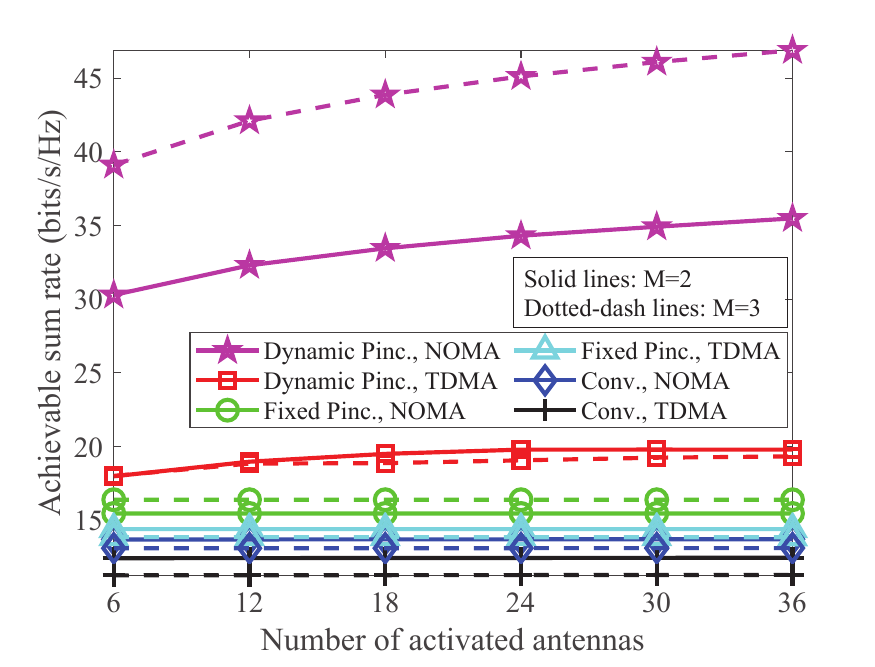}
		\caption{Achievable sum rate versus number of antennas.}
		\label{Fig_1WGmU_NumAntenna}
	\end{minipage}
\end{figure*}

Fig. {\ref{Fig_1WG1U_NumAntenna}} and Fig. \ref{Fig_1WG1U_Power} illustrate the impact of the number of activated antennas and total transmission power, respectively, where ``Dynamic Pinc." represents the  scheme of dynamic pinching antennas proposed in this paper, ``Fixed Pinc." represents the fixed-location pinching antennas deployed near the user with a  $\frac{\lambda}{2}$ antenna spacing, and ``Conv." represents the conventional uniform linear array deployed on the BS.
For both ``Fixed Pinc." and ``Conv.", we consider a performance upper bound with phase matching \cite{arXiv.2412.02376}. 
It can be observed from Fig. {\ref{Fig_1WG1U_NumAntenna}} that, except for the ``Dynamic Pinc." scheme with a carrier frequency of 5 GHz, the achievable rates of the other schemes all show an increasing trend with the number of activated antennas.
Due to the flexible deployment of pinching antennas, the performance gain becomes even greater as the number of antennas increases. 
However, when the carrier frequency is 5 GHz, the antenna spacing needs to be large to avoid coupling, resulting in a non-negligible increase to the distance between the outermost antennas relative to the distance to the user, especially when the number of antennas is large. Consequently, it exhibits a trend of first increasing and then decreasing. 
Fig. \ref{Fig_1WG1U_Power} demonstrates that the achievable rate performance for a single user improves as the total transmission power increases.
Furthermore, at the same carrier frequency, the proposed ``Dynamic Pinc." scheme always outperforms both ``Fixed Pinc." and ``Conv.", confirming the superiority of pinching antenna schemes.

\subsection{The SWMU case}

The $m$-th user is located at ${\psi}_{m} = \left(10\left(m-\frac{1}{2}\right), 10\left(m-\frac{1}{2}\right)-1, 0\right)$, the corresponding $m$-th sampling range is ${\cal{D}}_m =  \left\{\left[10\left(m-1\right), 10m\right],  \left[0, 10\right]\right\}$, and the feed point of the waveguide is $\left(0, 5, 3\right)$. 
We use the achievable sum rate as the metric to evaluate the performance of the proposed NOMA-based multi-user PASS. 
As shown in Fig. \ref{Fig_1WGmU_NumAntenna}, the NOMA-based pinching antenna solution demonstrates superior performance, proving the advantages of NOMA over TDMA, and the superiority of ``Dynamic Pinc." compared to ``Fixed Pinc." and ``Conv.". 
We observe that for the two schemes, ``Dynamic Pinc., NOMA" and ``Fixed Pinc., NOMA", their sum rates for the scenario with $M = 3$ are greater than those for the scenario with $M = 2$. 
The distances between users are relatively far apart, and each user has antennas close to them for ``Dynamic Pinc." and ``Fixed Pinc." schemes, resulting in a stronger received signal strength. 
For other schemes, as the number of users increases, the interference among users becomes greater or the available time slots per user decrease, leading to a smaller sum rate when $M = 3$ compared to when $M = 2$.

\begin{figure*}[t]
	\centering
	\begin{minipage}[t]{0.32\textwidth}
		\centering
		\includegraphics[width=6.0cm]{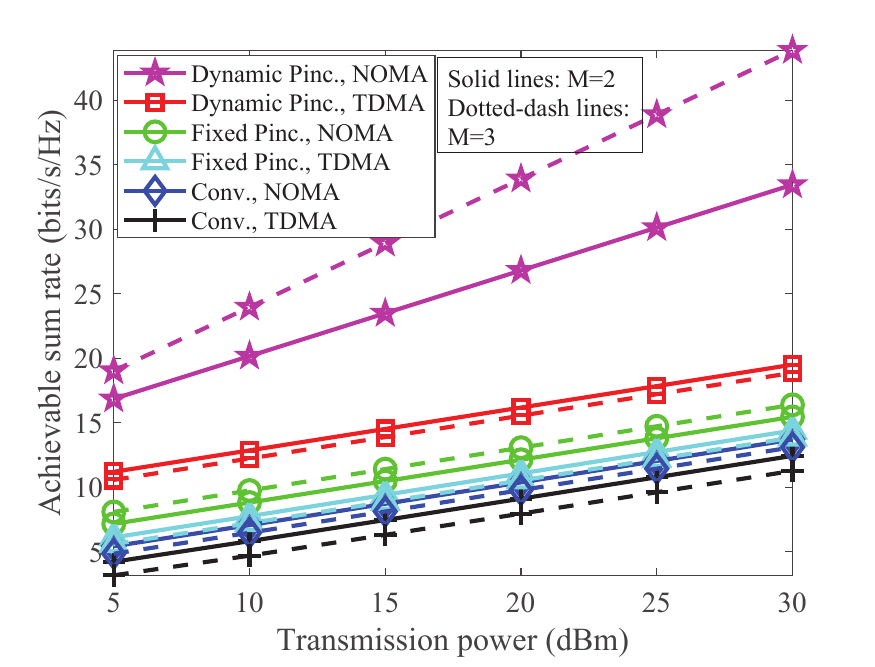}
		\caption{Achievable sum rate versus transmission power.}
		\label{Fig_1WGmU_Power}
	\end{minipage}
	\begin{minipage}[t]{0.32\textwidth}
		\centering
		\includegraphics[width=6.0cm]{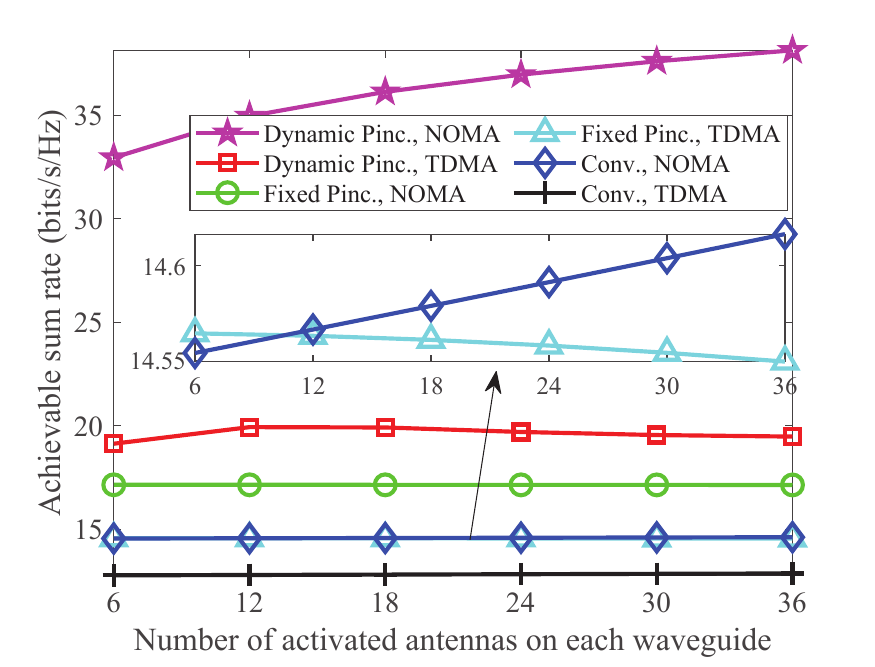}
		\caption{Achievable sum rate versus number of antennas.}
		\label{Fig_2WG2U_NumAntenna}
	\end{minipage}
	\begin{minipage}[t]{0.32\textwidth}
		\centering
		\includegraphics[width=6.0cm]{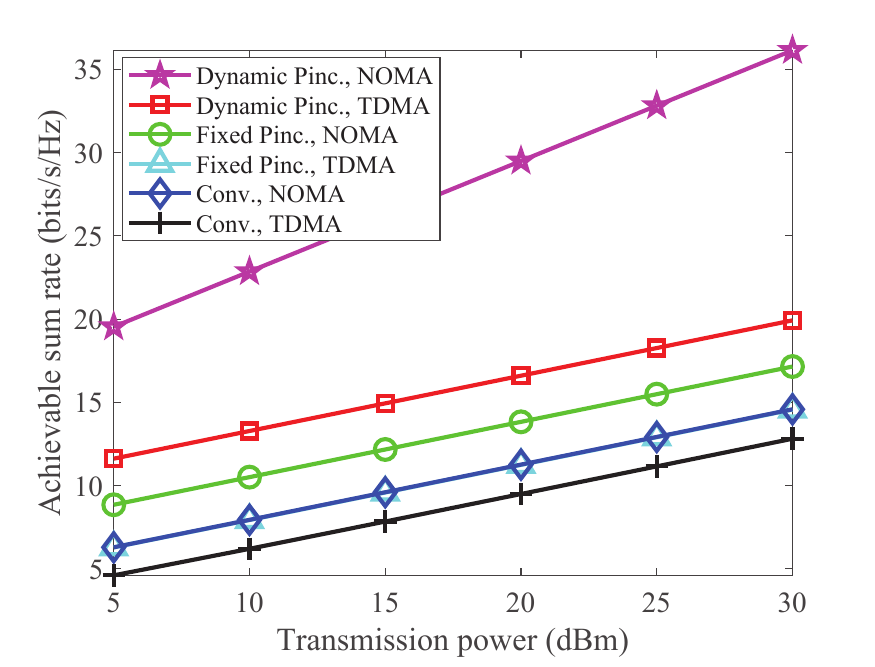}
		\caption{Achievable sum rate versus transmission power.}
		\label{Fig_2WG2U_Power}
	\end{minipage}
\end{figure*}

Fig. {\ref{Fig_1WGmU_Power}} evaluates the impact of the total transmission power on system performance, where all schemes have shown an upward trend.
As the power increases, the achievable sum rate of the pinching antenna scheme grows at a fastest rate. This is because the antennas in this scheme can be flexibly deployed, offering greater degrees of freedom and enabling better exploitation of the performance gains brought about by power enhancement. 

\subsection{The MWMU case}

The number of waveguides and the number of users are set as $Q = M = 2$.
The $m$-th user located at ${\psi}_{m} = \left(10\left(m-\frac{1}{2}\right), 10\left(m-\frac{1}{2}\right)-1, 0\right)$, the corresponding $m$-th sampling range is ${\cal{D}}_m =  \left\{\left[0, 10\right], \left[10\left(m-1\right), 10m\right]\right\}$, and the feed point of the $q$-th waveguide is $\left(0, 10\left(q-\frac{1}{2}\right), 3\right)$.  
As shown in Fig. {\ref{Fig_2WG2U_NumAntenna}} and Fig. {\ref{Fig_2WG2U_Power}}, as the number of activated pinching antennas or the transmission power increases, the achievable sum rate of the ``Dynamic Pinc., NOMA" scheme exhibits a notable increase and remains optimal compared to other schemes, confirming the effectiveness of our proposed solution. 
For other schemes, the increase in the number of antennas does not significantly enhance performance, as it is difficult for these schemes to exploit the gains provided by the flexibility as pinching antennas. 
Additionally, by comparing the scenario with M=2 in Fig. {\ref{Fig_1WGmU_NumAntenna}} and Fig. {\ref{Fig_1WGmU_Power}}, we find that the overall performance of the multiple waveguide PASS is better. 
On the one hand, multiple waveguides provide an additional dimension of freedom in antenna deployment. On the other hand, for each user, the large-scale pathloss to the antennas becomes smaller. 
Therefore, the performance of a multi-waveguide PASS is superior to that of a single-waveguide PASS.

\section{Conclusions}

In this paper, we investigated the beam training design problems for SWSU-PASS, SWMU-PASS, and MWMU-PASS. For the SWSU-PASS, we designed a scalable codebook, based on which we proposed a 3SBT scheme. 
For the SWMU-PASS, based on the scalable codebook, we proposed an improved 3SBT scheme.
For the MWMU-PASS, we presented a generalized expression of the received signal based on the partially-connected hybrid beamforming structure. 
Subsequently, we introduced an increased-dimensional scalable codebook design and proposed an increased-dimensional 3SBT scheme.
Numerical results revealed that the proposed beam training scheme significantly reduced the training overhead while maintaining reasonable rate performance compared to the 2D exhaustive search.
Furthermore, the proposed pinching-antenna scheme yielded better flexibility and improved system performance compared to fixed-location pinching antennas and  conventional array antennas.
Additionally, we noted that the proposed scheme exhibits better phase alignment performance at higher carrier frequencies.

\appendices

\section{Proof of Lemma \ref{MultipleWaveguide_y}} \label{AppendixB}

Consider the first case, where all the $N$ activated pinching antennas are deployed on the $i$-th waveguide and cluster around $\psi^{\rm{pin}}_{i,{\rm{U}}} = \left(x_{\rm{U}}, y_i^{\rm{wg}}, d\right)$, with $\psi^{\rm{pin}}_{i,{\rm{U}}} $ being the location on the $i$-th waveguide that is closest to the user. 
Denote the location of the $n$-th antenna on the $i$-th waveguide by $\tilde\psi_{i,n}^{\rm{pin}}$.
The channel between the activated antennas and the user is
\begin{equation}
	\begin{array}{l}
	{{\bf{h}}}_1 = \left[ \frac{\sqrt{\eta}e^{-2\pi j \left( \frac{1}{\lambda }\left| \psi_{\rm{U}} -\tilde\psi_{i,1}^{\rm{pin}}\right| + \frac{1}{\lambda_g }\left| \psi_0^{\rm{pin}} -\tilde\psi_{i,1}^{\rm{pin}}\right| \right)}}{\left | \psi_{\rm{U}} -\tilde\psi_{i,1}^{\rm{pin}}\right |} \right.  \vspace{1ex} \\
	\;\;\;\;\;\;\;\; \cdots \left. \frac{\sqrt{\eta}e^{-2\pi j \left( \frac{1}{\lambda }\left| \psi_{\rm{U}} -\tilde\psi_{i,N}^{\rm{pin}}\right| + \frac{1}{\lambda_g }\left| \psi_0^{\rm{pin}} -\tilde\psi_{i,N}^{\rm{pin}}\right| \right)}}{\left | \psi_{\rm{U}} -\tilde\psi_{i,N}^{\rm{pin}}\right |} \right]^T.
	\end{array}
\end{equation}
We note that a movement of a few wavelengths has a limited impact on distance $\left| \psi_{\rm{U}} -\tilde\psi_{i,n}^{\rm{pin}} \right|$, 
i.e., $\left| \psi_{\rm{U}} -\tilde\psi_{i,n}^{\rm{pin}} \right| \approx \left| \psi_{\rm{U}} -\psi_{i,{\rm{U}}}^{\rm{pin}} \right|$, $\forall n = 1, \cdots, N$. 
Therefore, the received signal strength of the user is
\begin{equation}
	\begin{array}{l}
		\left|r_1\right| = {\frac{P}{N}} \left|\sum\limits_{n = 1}^{N} {\frac{{\sqrt{\eta} e^{-2\pi j \left( \frac{1}{\lambda }\left| \psi_{\rm{U}} -\tilde\psi_{i,n}^{\rm{pin}}\right| + \frac{1}{\lambda_g }\left| \psi_0^{\rm{pin}} -\tilde\psi_{i,n}^{\rm{pin}}\right| \right)}}}{{\left| \psi_{\rm{U}} -\tilde\psi_{i,n}^{\rm{pin}}  \right| }}} \right|^2 \vspace{1ex}\\
		\;\;\;\;\;\; \overset{\left( b \right)}{=} {\frac{\eta P}{N}} \left|\sum\limits_{n = 1}^{N} {\frac{{1}}{{\left| \psi_{\rm{U}} -\tilde\psi_{i,n}^{\rm{pin}}  \right| }}} \right|^2  \overset{\left( c \right)}{\approx} {\frac{\eta P}{N}} \left|\sum\limits_{n = 1}^{N} {\frac{{1}}{\sqrt{ \left({y_i^{\rm{wg}}} - y_{\rm{U}}\right)^2 + d^2 }}}  \right|^2,
	\end{array} 
\end{equation}
where $\left(b\right)$ the design of the antenna locations can ensure that $\mod \left\{\frac{2\pi}{\lambda }\left| \psi_{\rm{U}} -\tilde\psi_{i,n}^{\rm{pin}}\right| + \frac{2\pi}{\lambda_g }\left| \psi_0^{\rm{pin}} -\tilde\psi_{i,n}^{\rm{pin}}\right| , 2\pi\right\}=0$, and $\left(c\right)$ is due to the fact that the principle for determining antenna locations is to cluster around $\psi^{\rm{pin}}_{i,{\rm{U}}}$ within a few wavelengths, and therefore the difference in the $x$-coordinate between $\psi_{\rm{U}}$ and $\psi_{i,{\rm{U}}}^{\rm{pin}}$ can be neglected for the calculation of distance $\left| \psi_{\rm{U}} -\psi_{i,{\rm{U}}}^{\rm{pin}} \right|$.

Consider another case, where $N_i$ activated pinching antennas are deployed on the $i$-th waveguide and $N_j$ activated pinching antennas are deployed on other waveguides, satisfying $0 \le N_i \le N$, $0 \le N_j \le N$, $N_i + N_j = N$ and $i \ne j$. 
The channel between the $\left\langle {q,n} \right\rangle$-th antenna and the user is 
\begin{equation}
	\begin{array}{l}
	h_{q,n}=\frac{\sqrt{\eta}e^{-2\pi j \left( \frac{1}{\lambda }\left| \psi_{\rm{U}} -\tilde\psi_{q,n}^{\rm{pin}}\right| + \frac{1}{\lambda_g }\left| \psi_0^{\rm{pin}} -\tilde\psi_{q,n}^{\rm{pin}}\right| \right)}}{\left | \psi_{\rm{U}} -\tilde\psi_{q,n}^{\rm{pin}}\right |}.
	\end{array}
\end{equation}
The channel between the activated antennas and the user is
\begin{equation}
	{\bf{h}}_2 = \left[h_{i,1}, \cdots, h_{i,N_i}, h_{j,1}, \cdots, h_{j,N_j}\right]^T.
\end{equation}
Still, consider that the principle for determining antenna locations is to cluster around $\psi^{\rm{pin}}_{q,{\rm{U}}}$, where $\psi^{\rm{pin}}_{q,{\rm{U}}} $ is the location on the $q$-th waveguide that is closest to the user. 
Denote the location of the $n$-th pinching antenna on the $q$-th waveguide by $\tilde\psi_{q,n}^{\rm{pin}}$.
Then we have $\left| \psi_{\rm{U}} -\tilde\psi_{q,n}^{\rm{pin}} \right| \approx \left| \psi_{\rm{U}} -\psi_{q,{\rm{U}}}^{\rm{pin}} \right|$. 
Therefore, the signal strength received by the user from the $N_i +N_j$ antennas can be expressed as
\begin{equation}
	\begin{array}{l}
		\left|r_2\right| \overset{\left( b \right)}{=} {\frac{\eta P}{N}} \left|\sum\limits_{n = 1}^{N_i} {\frac{{1}}{{\left| \psi_{\rm{U}} -\tilde\psi_{i,n}^{\rm{pin}}  \right| }}} + \sum\limits_{n = 1}^{N_j} {\frac{{1}}{{\left| \psi_{\rm{U}} -\tilde\psi_{j,n}^{\rm{pin}}  \right| }}}\right|^2 \vspace{1ex}\\
		\;\;\;\;\;\; \overset{\left( c \right)}{\approx} {\frac{\eta P}{N}} \left|\sum\limits_{n = 1}^{N_i} {\frac{{1}}{\sqrt{ \left({y_i^{\rm{wg}}} - y_{\rm{U}}\right)^2 + d^2 }}} + \sum\limits_{n = 1}^{N_j} {\frac{{1}}{\sqrt{\left({y_j^{\rm{wg}}} - y_{\rm{U}}\right)^2 + d^2 }}} \right|^2\vspace{1ex} \\
		\;\;\;\;\;\; \overset{\left( d \right)}{\le}  {\frac{\eta P}{N}} \left|\sum\limits_{n = 1}^{N} {\frac{{1}}{\sqrt{ \left({y_i^{\rm{wg}}} - y_{\rm{U}}\right)^2 + d^2 }}}  \right|^2 = \left|r_1\right|,
	\end{array} 
\end{equation}
where $\left(d\right)$ is due to the condition $\left| y_i^{\rm{wg}} - y_{\rm{U}}\right| < \left| y_j^{\rm{wg}} - y_{\rm{U}} \right|$, i.e., $\left(y_i^{\rm{wg}} - y_{\rm{U}}\right)^2< \left( y_j^{\rm{wg}} - y_{\rm{U}} \right) ^2$, $\forall$ $i, j = 1, \cdots, Q$, $i \ne j$.

Furthermore, for other cases that the locations of activated antennas do not satisfy the minimum distance principle, the distances between these antennas and the user will be larger, which implies that the signal strength received by the user will be weaker. 
Through the above analysis, it is proven that $\left|r_1\right|$ is the largest received signal strength, completing the proof of {\textit{Lemma \ref{MultipleWaveguide_y}}}.

\bibliographystyle{IEEEtran}
\bibliography{ref/ref_PinchingAntenna}


\end{document}